\documentclass[journal,twoside,web]{ieeecolor}
\usepackage{tmi}
\usepackage{cite}
\usepackage{amsmath,amssymb,amsfonts}
\usepackage{algorithmic}
\usepackage{graphicx}
\usepackage{textcomp}

\usepackage[ruled,noline]{algorithm2e}
\usepackage{algorithmic}

\usepackage{cite}
\usepackage{amsmath,amssymb,amsfonts}
\usepackage{algorithmic}
\usepackage{graphicx}
\usepackage{textcomp}
\usepackage{amsmath}
\usepackage{comment}

\usepackage{graphicx}
\usepackage{multicol}
\usepackage{multirow}

\definecolor{mycolor}{rgb}{0.858, 0.188, 0.478}
\definecolor{blue}{rgb}{0, 0, 1}
\def\BibTeX{{\rm B\kern-.05em{\sc i\kern-.025em b}\kern-.08em
    T\kern-.1667em\lower.7ex\hbox{E}\kern-.125emX}}
\begin{document}
\title{One shot PACS: \underline{P}atient specific \underline{A}natomic \underline{C}ontext and \underline{S}hape prior aware recurrent registration-segmentation of longitudinal thoracic cone beam CTs}
\author{Jue Jiang and Harini Veeraraghavan
\thanks{The manuscript has been submitted on May 4$^{th}$,2021 for review. This work was supported by the MSK Cancer Center core grant P30 CA008748.}
\thanks{Jue Jiang and Harini Veeraraghavan are with the Department of Medical Physics, Memorial Sloan Kettering Cancer Center, NY, New York, USA.(e-mail: jiangj1@mskcc.org and veerarah@mskcc.org).}}

\maketitle

\begin{abstract}
Image-guided adaptive lung radiotherapy requires accurate tumor and organs segmentation from during treatment cone-beam CT (CBCT) images. Thoracic CBCTs are hard to segment because of low soft-tissue contrast, imaging artifacts, respiratory motion, and large treatment induced intra-thoracic anatomic changes. Hence, we developed a novel Patient-specific Anatomic Context and Shape prior or PACS-aware 3D recurrent registration-segmentation network for longitudinal thoracic CBCT segmentation. Segmentation and registration networks were concurrently trained in an end-to-end framework and implemented with convolutional long-short term memory models. The registration network was trained in an unsupervised manner using pairs of planning CT (pCT) and CBCT images and produced a progressively deformed sequence of images. The segmentation network was optimized in a one-shot setting by combining progressively deformed pCT (anatomic context) and pCT delineations (shape context) with CBCT images. Our method, one-shot PACS was significantly more accurate (p$<$0.001) for tumor (DSC of 0.83 $\pm$ 0.08, surface DSC [sDSC] of 0.97 $\pm$ 0.06, and Hausdorff distance at $95^{th}$ percentile [HD95] of 3.97$\pm$3.02mm) and the esophagus (DSC of 0.78 $\pm$ 0.13, sDSC of 0.90$\pm$0.14, HD95 of 3.22$\pm$2.02) segmentation than multiple methods. Ablation tests and comparative experiments were also done. 
\end{abstract}


\section{Introduction }
\label{sec:introduction}
Adaptive image-guided radiation treatments (AIGRT) of lung cancers require accurate segmentation of tumor and organs at risk (OAR) such as the esophagus from during treatment cone-beam CT (CBCT)\cite{sonke2019}. Tumors are difficult to segment due to very low soft-tissue contrast on CBCT, imaging artifacts, radiotherapy (RT) induced radiographic and size changes, and large intra- and inter-fraction motion\cite{sonke2019}. Normal organ like the esophagus is also hard to segment due to low soft-tissue contrast and displacements exceeding 4mm between treatment fractions\cite{wangRadOnc2013}. 
\par
Cross-modality deep learning methods have used structure constraints from planning CT (pCT)\cite{kuckertz2020}, as well as MRI contrast as prior knowledge to improve pelvic organ \cite{Fu2020MedPhys} and lung tumor\cite{jiang2021CBCTMedPhys} segmentation from CBCT. However, expert delineated CBCTs needed for training are not routinely segmented and suffer from high inter-rater variability \cite{Altorjai2012CBCT}. 
\par 
Atlas-based image registration methods overcome the issue of limited segmented datasets by directly propagating segmentations\cite{dong2017Neurocomputing} as well as by providing synthesized images as augmented data for segmentation training\cite{zhaoCVPR2019,he2020deep,wang2020lt}. Cross-domain adaptation based synthetic CBCT generation based data augmentation \cite{jia2019MICCAI} is another promising approach used for pelvic organs segmentation. \textcolor{black}{A hybrid approach\cite{ZhouMIA2021} combining data augmentation using cross-domain adaptation of pCT, MRI, and CBCT with multi-modality registration was used for liver segmentation from during treatment CBCTs.} 
\par
Multi-task networks\cite{xu2019deepatlas, he2020deep, estienne2019u,beljaardsCrossStitch2020} handle limited datasets by using implicit data augmentation available from the different tasks through the losses to jointly optimize registration and segmentation. Notably, these  methods have shown feasibility for one and few-shot normal organ segmentation\cite{xu2019deepatlas,he2020deep,cui2020unified}, using CT-to-CT or MRI-to-MRI registration. \textcolor{black}{Planning CT to CBCT registration is harder because of low soft-tissue contrast and narrow field of view (FOV) on CBCT}\cite{ZhouMIA2021}.
\par
Prior works applied to CBCT registration aligned large moving organs\cite{he2020deep,Fu2020MedPhys,ZhouMIA2021,Foote2019IPMI,Kearney2018}. \textcolor{black}{We tackle more challenging lung tumor segmentation from CBCT for longitudinal response assessment in tumors undergoing treatment}. Previously, longitudinal tracking in anatomy depicting large changes, such as growing infant brains\cite{dong2017Neurocomputing} and pre- and post-surgical brains\cite{Chitphakdithai2010MICCAI} aligned same modalities with high contrast (MRI-to-MRI). \textcolor{black}{Cross-domain adaptation based synthetic CT\cite{fu2020cone} as well as deep network combined with scale invariant feature transform (SIFT) detected features\cite{Kearney2018}, and surface points registration of multiple modalities\cite{ZhouMIA2021} are example approaches used to handle low soft-tissue contrast on CBCT.} We address multiple challenges, namely, multi-modal registration (pCT to CBCT), longitudinal registration of highly deforming diseased and healthy tissues with altered appearance from treatment, and segmentation on low soft-tissue contrast CBCT.
\par 
In order to tackle all these challenges, our approach combines an end-to-end trained joint recurrent registration network (RRN) and recurrent segmentation network (RSN).  \textcolor{black}{Our approach models large local deformations by computing progressive deformations that incrementally improves regional alignment. This is accomplished by using convolutional long-short term memory (CLSTM)\cite{shi2015convolutional} to implement the recurrent units of RRN and RSN. CLSTM models long-range temporal deformation dynamics, needed to model the progressive deformations in regions undergoing large deformations. The convolutional layers used in CLSTM models the spatial dynamics of a dense 3D flow field compared to 1D information computed by LSTM. Our approach increases flexibility to capture longitudinal size and shape changes in tumors compared to the Recurrent Registration Neural Network (R2N2)\cite{sandkuhler2019recurrent}, which computes parameterized local deformations. Finally, in order to handle low contrast on CBCT, RSN combines progressively aligned anatomical context (pCT) and shape (pCT delineation) prior produced by a jointly trained RRN. Hence, we call our approach patient-specific anatomic context and shape prior or PACS-aware registration-segmentation network. We show that our approach is more accurate than multiple methods.}
\par
The RRN is trained in an unsupervised way using only pairs of target and moving images \textcolor{black}{without structure guidance from segmented pCT or CBCT. RSN is optimized with a single segmented CBCT example and combines progressively warped pCT images and delineations produced by the RRN with CBCT for one-shot training.} Our contributions are: 
%
\begin{itemize}
    \item Multi-modal recurrent joint registration-segmentation approach to handle large anatomic changes in tumors undergoing treatment and highly deforming esophagus. 
    \item One-shot segmentation with patient-specific anatomic context and shape priors, which handles tumor segmentation despite varying size and locations. To our best knowledge, this is the first one-shot learning approach for longitudinal segmentation of lung tumors from CBCT. 
    \item A recurrent registration network that interpolates dense flow field using only a pair of pCT and CBCT images and optimized with unsupervised training. 
    \item Comprehensive comparison, ablation, and network design experiments to study accuracy. 
\end{itemize}

	\begin{figure*}[t]
		\begin{center}
			\includegraphics[width=1.9\columnwidth,scale=0.1]{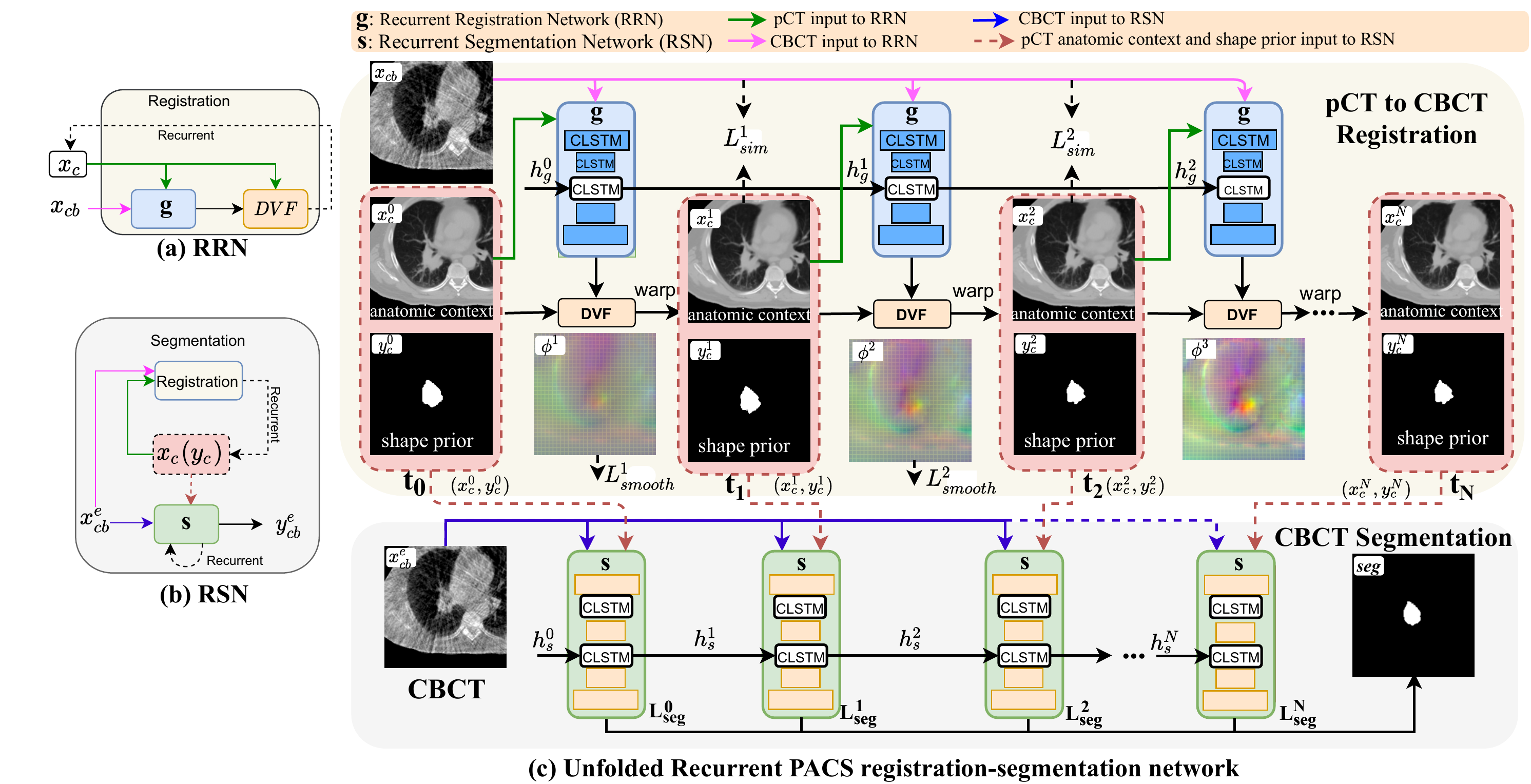}
			\vspace{-0.05cm}\setlength{\belowcaptionskip}{-0.4cm}\setlength{\abovecaptionskip}{0.08cm}\caption{\small One-shot PACS registration-segmentation network. RRN $g$ uses N CLSTM steps to align $x_c$ to $x_{cb}$. RSN $s$ uses N+1 CLSTM steps to segment $x_{cb}$, where $\{x_{c}^{t}, y_{c}^{t}\}$, \textcolor{black}{$1 \leq t \leq N$} produced by RRN are used in the RSN to provide spatial and anatomic priors. 
			 \label{fig:overview_method}}
		\end{center}
	\end{figure*}
\section{Related works}

\subsection{Medical image registration-based segmentation} 
\textcolor{black}{One-shot and few-shot learning strategies extract a model from only a single or a few labeled examples.} Hence, these are attractive options for medical image analysis where large number of expert segmented cases are not available. As elucidated by Wang et.al\cite{wang2021alternative}, a key difference between few shot learning applied to natural vs. medical images is that, in the former, learning is concerned with extracting a model to recognize a new class based on appearance similarities to previously learned classes. Medical image analysis methods are concerned with better modeling the anatomic similarity between subjects using few examples where all classes are available. The challenge is to extract a representation that is robust to imaging and anatomic variability among patients. 
\par
Iterative registration methods sidestep the issue of learning by using the available segmented cases as atlases\cite{avants2008symmetric}. Assuming that the atlases represent the variability in the patient anatomy, these methods provide reasonably accurate tissue segmentations. Unsupervised deep learning-based synthesis of realistic training samples\cite{zhaoCVPR2019} as augmented datasets are a more robust option, because they do not suffer from catastrophic failures. Also, once trained the network produces computationally fast segmentations unlike iterative registration. Nonetheless, accuracy is reduced due to poor image quality and large anatomical changes\cite{brockMedPhys2017}.
\par
Joint registration-segmentation methods\cite{estienne2019u,ElmahdyMedPhys2019,beljaardsCrossStitch2020,xu2019deepatlas} are more accurate than registration-based segmentation. This is because, these methods model the interaction between registration and segmentation features and improve accuracy. 
Furthermore, these approaches are amenable to training with few segmented examples including semi-supervised learning. For example, a registration network was used to segment unlabeled data for training\cite{wang2020lt,xu2019deepatlas} as well as create augmented samples through random perturbations in the warped images\cite{he2020deep}. These methods benefit by using the segmentation network to provide additional regularization losses to optimize the registration network training such as through segmentation consistency\cite{xu2019deepatlas} and cycle consistency losses\cite{wang2020lt}. However, the one-step registration computed by these methods may not handle very large deformations. 
\par
Multi-stage \cite{deVos2019MedIA} and recursive cascade registration\cite{zhao2019recursive} methods have shown that incrementally refined outputs produced from intermediate steps as inputs to subsequent steps increases accuracy to model large deformations but require large number of parameters. R2N2\cite{sandkuhler2019recurrent} improves on these methods by using gated recurrent units with local Gaussian basis functions and captures organ deformations occurring in a breathing cycle from MRI.
 \subsection{Shape and spatial priors for registration-segmentation}
  Template shape constraints\cite{lee2019tetris} as well as population level anatomical priors learned using a generative model were previously used to regularize same modality registrations. Segmentation as auxiliary supervised information has shown to provide more accurate registrations\cite{balakrishnan2019voxelmorph,zhang2011estimating,xu2019deepatlas,lee2019image}. Structure guidance from pCT\cite{kuckertz2020} and anatomical priors priors have shown to improve accuracy. Improving on prior works that used either shape\cite{kuckertz2020,zhang2011estimating,balakrishnan2019voxelmorph,lee2019image,lee2019tetris} or anatomical priors, we use both priors and show accuracy gains, even in the one-shot segmentation training scenario, where a single segmented CBCT example is used for training. \textcolor{black}{Our joint recurrent registration-segmentation network computes a dense 3D flow field but uses fewer parameters than cascaded methods\cite{zhao2019recursive,deVos2019MedIA} and is more accurate than  R2N2\cite{sandkuhler2019recurrent}.}
\par
 \textbf{Rationale for combining pCT as anatomical context and and delineations as shape prior: \/}\rm  The pCT has a higher soft-tissue contrast than the CBCT scans, which can provide a spatially aligned anatomic context to improve inference from lower-contrast CBCT. We also expect that the pCT segmentations used as patient-specific priors to segment CBCT will be informative of the tumor and organ shape for segmentation. 
\newpage
\section{Method}
\subsection{Background}

\textbf{Problem setting and approach: \/} \textcolor{black}{Given a single segmented CBCT example $\{x_{cb}^{e},y_{cb}^{e}\}$, it's corresponding delineated pCT $\{x_c^{e}, y_c^{e}\}$, as well as several unsegmented CBCTs $x_{cb} \in X_{CBCT}$ and their corresponding segmented pCTs $\{x_c, y_c\} \in \{X_{C},Y_{C}\}$, our goal is to construct a model to generate  tumor and esophagus segmentations from weekly CBCTs.} 
\par
Our approach uses a joint recurrent registration-segmentation network (Fig.~\ref{fig:overview_method}). \textcolor{black}{The recurrent registration network (RRN) $g$ and recurrent segmentation network (RSN) $s$ are implemented using CLSTM\cite{shi2015convolutional}}. \textcolor{black}{RRN aligns $x_{c}$ to $x_{cb}$ and produces progressively deforming $\{x_c^t,y_c^t\}$ in $N$ CLSTM steps, where $1 \leq t \leq N$. RSN computes a segmentation $y_{cb}$ for $x_{cb}$ by combining progressively warped $\{x_{c}^{t}, y_{c}^{t}\}$ produced by the RRN using $N+1$ CLSTM steps. As shown in Fig.~\ref{fig:overview_method}, the first CLSTM step ($t=0$) of RSN uses input pCT and it's delineation ($x_{c}^{0}, y_{c}^{0}$) with $x_{cb}$.} \textcolor{black}{On the other hand, the CLSTM steps $t \geq 1 $ of RSN use the outputs of RRN ($x_{c}^{t}, y_{c}^{t}$) with $x_{cb}$.} 

\textcolor{black}{\textbf{Classical dynamic system vs. basic recurrent network vs. LSTM vs. CLSTM}
A classical dynamic system (CDS) uses shared feature layers to produce outputs sequentially\cite{Goodfellow_book}, represented as, $x^{t} = f(x^{t-1}; \theta)$. A basic recurrent neural network (RNN) also includes a hidden state\cite{Goodfellow_book}, ($x^{t} = f(x^{t-1}, h^{t-1}; \theta)$) in order to blend information from preceding temporal step. LSTM and CLSTM are a type of RNN, which use feedback through forget gate and memory cells to capture long range temporal information. CLSTM uses convolution layers to model dense spatial dynamics whereas LSTM models 1D dynamics through fully connected layers\cite{shi2015convolutional}. \textcolor{black}{A diagram of the differences are shown in 
Supplementary Fig 2.}}

\textbf{Convolutional long short term memory network: \/}\rm CLSTM is a recurrent network that was introduced to model the dynamics within 2D spatial region\cite{shi2015convolutional} \textcolor{black}{via convolution. We extended CLSTM to model deformation dynamics in a 3D spatial region. Moreover, our approach models the large deformation dynamics as an interpolated temporal deformation sequence (for RRN) or an interpolated segmentation sequence (for RSN), given only the start (pCT and its contour) and end images (CBCT image to be segmented) of the sequence.} 

A CLSTM unit is composed of a memory cell $c^{t}$, which accumulates the state $x^{t}$ at step $t$, a forget gate $f^{t}$ that keeps track of relevant state information from the past, a hidden state $h^{t}$, which encodes the state, as well as input state $i^{t}$, and output gate $o^{t}$. The CLSTM components are updated as: 
\begin{equation}
\begin{split}
f^{t}&=\sigma(W_{xf}*x^{t}+W_{hf}*h^{t-1}+b_{f})\\
i^{t}&=\sigma(W_{xi}*x^{t}+W_{hi}*h^{t-1}+b_{i})\\
\tilde{c}^{t}&=tanh(W_{x\tilde{c}}*x^{t}+W_{h\tilde{c}}*h^{t-1}+b_{\tilde{c}})\\
o^{t} &= \sigma(W_{xo}*x^{t}+W_{ho}*h^{t-1}+b_{o})\\
c^{t} &= f^{t} \odot c^{t-1}+i^{t}\odot \tilde{c}^t\\
h^{t} &= o^{t} \odot tanh(c^{t}),
\end{split}
\end{equation}
where, $\sigma$ is the sigmoid activation function, $*$ the convolution operator, $\odot$ the Hadamard product, and $W$ the weight matrix. 
\subsection{Planning CT to CBCT deformable image registration}
The RRN $g$ computes a deformation of $x_c$ to $x_{cb}$, expressed as $g(x_c, x_{cb}): \theta_{g}(x_c) \rightarrow x_{c}^{cb}$ as a sequence of deformation vector fields (DVF) using $N$ steps: $\phi_{c}^{cb} = \phi^{1}  \circ \phi^{2}   ... \circ \phi^{N}$. $\phi^{t} : I + u^{t}$, where I is the identity and $u$ is the displacement vector field to deform pixels from a currently warped pCT $x_{c}^{t} \in \mathbb{R}^{L \times Q \times P}$ closer to the coordinates of the CBCT image as $x_{c}^{t+1}$. \textcolor{black}{No structure guidance from segmented pCT or CBCT is used and the RRN is optimized in an unsupervised manner using pairs of pCT and CBCT images $\{x_{c}, x_{cb}\}$}.
\par
\textcolor{black}{The inputs to the RRN consist of channel-wise concatenated image pairs and hidden state, $\{x_{c}^{t-1}, x_{cb}, h_g^{t-1}\}$, where $h_{g}^{0}$ is initialized to 0 and $x_{c}^{0}$ = $x_{c}$ (Fig.~\ref{fig:overview_method}). The intermediate CLSTM steps $t \geq 1$ receive the hidden state $h_g^{t-1}$ produced from prior CLSTM step $t-1$. RRN at CLSTM step $t$ outputs a warped pCT image $x_{c}^{t}$ and the updated hidden state $h_{g}^{t}$.} With $x_{c}^{0}$ = $x_{c}$, it's delineation, $y_{c}^{0}$ = $y_{c}$, and $\phi^{t}$=g($x_{c}^{t-1}$,$x_{cb}$,$h_g^{t-1}$), the warped pCT and its delineation are computed as:
\begin{equation}
\setlength{\abovedisplayskip}{1pt}
\setlength{\belowdisplayskip}{1pt} 
\begin{split}
x_{c}^{t} = x_{c}^{t-1} \circ \phi^{t} \\
y_{c}^{t} = y_{c}^{t-1} \circ \phi^{t}.
\end{split}
\label{eqn:reg_phi1}
\end{equation}
\par 
RRN is optimized using image similarity $L_{sim}$ and smoothness loss $L_{smooth}$ measured from the flow field gradient, with a tradeoff parameter $\lambda_{smooth}$ to control image similarity and deformation smoothness. $L_{sim}$ is computed using Normalized Cross-Correlation (NCC) between the CBCT $x_{cb}$ and the $N$ warped pCTs $x_{c}^{t}$ produced by CLSTM steps. \textcolor{black}{NCC was computed locally using window of 5$\times$5$\times$5 centered on each voxel to improve robustness to CT and CBCT intensity differences\cite{balakrishnan2019voxelmorph}}. $L_{sim}$ was computed as:
\begin{equation}
\setlength{\abovedisplayskip}{1pt}
\setlength{\belowdisplayskip}{1pt} 
\begin{split}
L_{sim} = \textcolor{black}{-\underset{t=1}{\overset{N}\sum}\ L_{sim}^t}=-\underset{t=1}{\overset{N}\sum}\ NCC(x_{c}^{t},x_{cb})/N,
\end{split}
\end{equation}
\textcolor{black}{NCC$(x_{c}^{t},x_{cb})$ at each step $t$ is an average of all local NCC calculations.}
The smoothness loss term is computed as:
\begin{equation}
\setlength{\abovedisplayskip}{1pt}
\setlength{\belowdisplayskip}{1pt} 
\begin{split}
L_{smooth} = \textcolor{black}{\underset{t=1}{\overset{N}\sum}\ L_{smooth}^t}= \underset{t=1}{\overset{N}\sum}\ \underset{p \in \Omega}{\sum}\ ||\nabla \phi^t(p)||^{2}/N.   
\end{split}
\end{equation}

The total registration loss is computed as $L_{reg}$ = $L_{sim} + \lambda_{smooth} \times L_{smooth}$. 

\subsection{One shot Patient-specific anatomic context and shape prior-aware CBCT segmentation}

\textcolor{black}{The CLSTM step $t$ of RSN uses a channel-wise concatenated input $\{x_{c}^{t}, y_{c}^{t},x_{cb}, h_{s}^{t}\}$ to compute a segmentation $y_{cb}^{t} = s(x^{t}_{c}, y^{t}_{c}, x_{cb}, h^{t}_{s})$. $x_{c}^{t}$ and $y_{c}^{t}$ are produced by the RRN at CLSTM step $t$ and $h_s^t$ is the hidden state of $s$ from CLSTM at time t, 1} \textcolor{black}{$\leq$} $t$ $\leq$ $N$ (Eqn.~\ref{eqn:reg_phi1}). As shown in Fig.~\ref{fig:overview_method}, the first CLSTM step of RSN t $=$ 0 is initialized with a weak prior from undeformed pCT inputs ($x_{c}^0,y_{c}^0,x_{cb},h_s^0$). CBCT segmentation $y_{cb}$ is produced after $N+1$ CLSTM steps. 
\par
In one-shot training, only a single exemplar segmented CBCT $\{x_{cb}^{e},y_{cb}^{e}\}$ is available. RSN learns a mapping $\theta_{S}(.)$ of an image to it's segmentation, $s(x_c, y_c, x_{cb}^{e}): \theta_{s}(x_{cb}^{e}) \rightarrow y_{cb}^{e}$.  The segmentation loss is computed from segmentations computed in all CLSTM steps $0 \leq t \leq N$ of RSN as:
\begin{equation}
    \begin{split}
    \setlength{\abovedisplayskip}{1pt}
    \setlength{\belowdisplayskip}{1pt}
    L_{seg} =\underset{t=0}{\overset{N}\sum}L_{seg}^t=\underset{t=0}{\overset{N}\sum} logP(y_{cb}^{e}|s(x^{t}_c, y^{t}_c,x^{e}_{cb},h^{t}_{s})).
    \label{eqn:segmentation_loss2}
    \end{split}
\end{equation}
The losses, $L_{seg}^{0}$,...$L_{seg}^{N-1}$ provide deep supervision to train RSN. 

\subsubsection{Online Hard Example Mining (OHEM) Loss} 
\textcolor{black}{OHEM loss was previously used to improve stability of training in the presence of highly imbalanced classes in few-shot learning\cite{cui2020unified,wu2016highOHEM}. Training stability is improved because the pixels considered for gradient computation change based on model output, and which acts as a form of online bootstrapping.} OHEM loss focuses the network towards pixels that are hard to classify in a minibatch. Hard pixels are those that are associated with a small probability of producing the correct classification, or $ p_{m,c} < \tau$, \textcolor{black}{where $p_{m,c}$ is the probability associated to a class c for a pixel $m$, and $\tau$ is the probability threshold for selecting the hard pixels}.  We set $\tau = 0.7$ and $K = 10,000$, the minimum number of hard pixels to be used within each mini-batch.
The OHEM loss is computed as: 
\begin{equation}
    \begin{split}
    &L_{seg}^{ohem} 
    =\\&
    \underset{t=0}{\overset{N}\sum} \underset{m=1}{\overset{M}\sum} \underset{c=1}{\overset{C}\sum} 1\{p_{m,c}\textless\tau\}log P_{m,c}(y_{cb}^{e}|s(x^{t}_c, y^{t}_c,x^{e}_{cb},h^{t}_{s})),
    \label{eqn:segmentation_loss2_OHEM}
    \end{split}
\end{equation}
where $\tau$ $\in$ (0,1] is a threshold; 1$\{\star\}$ equals to one when the condition inside holds; C is the total class number; $M$ is the total number of voxels inside one mini-batch.
\subsubsection{Joint registration-segmentation network optimization}

\par 
The RRN network parameters are fixed when training the RSN and vice versa. The number of training examples available to optimize RRN $\mathcal{K} \gg 1$, where 1 is the number of examples available to optimize RSN. We replicated the one-shot exemplar 0.1$\times \mathcal{K}$ times through online augmentation to improve training stability. The number of replications was determined experimentally. This example replication is akin to upsampling (not to be confused with image upsampling) strategy used in machine learning for improving model generalizability. The training examples are shuffled to randomize the order of network updates. RRN $g$ is updated using the gradient $-\Delta_{\theta_{g}}(L_{reg})$. RSN $s$ is updated using the gradient $-\Delta_{\theta_{s}}(L_{seg}^{ohem})$. The detailed training procedure for one-shot PACS registration-segmention method is in Algorithm \ref{algo:algorithm}.


\begin{algorithm}[t]
	\footnotesize
	\LinesNumbered
	{\SetKwInOut{Input}{input}\SetKwInOut{Output}{output}
	 \Input{Unlabeled CT and CBCT dataset ($x_c,x_{cb}$) $\in$ $\{X_{c},X_{cb}\}\textcolor{black}{^\mathcal{K}}$, Exemplar segmented CBCT and corresponding segmented pCT, ($\{x_{cb}^e,y_{cb}^e, x_{c},y_{c}\}$)$^{1}$ $\in$ $\{X_{c}, Y_{c}, X_{cb}, Y_{cb}\}^{\mathcal{K}}$,$\mathcal{K}$ is the number of pCT, CBCT pairs, 1 refers to a single example.}
	 \Output{Registration model $\theta_{g}$ to align $x_c$ to $x_{cb}$ and segmentation model $\theta_{s}$  to segment $x_{cb}$}
	 Upsample one-shot CBCT example ($x_{cb}^e,y_{cb}^e$) and its corresponding pCT ($x_{c},y_{c}$) to produce $\mathcal{K}/10$ examples\;
	 $\theta_{g}$,$\theta_{s}$ $\gets$ initialize \;
	 \For {Epoch id $\leq$ Maximum Epoch}{
	 \For {Iter $\leq$ Maximum Iter} {

	    \eIf { Registration Flag}{
            
            $x_{c}$, $x_{cb}$ $\gets$ sample mini-batch from $\{X_{c},X_{cb}\}$\;
		$L_{reg}$ $\gets$ calculated using (3) (4)\;
		$\theta_{g} \overset{+}{\gets} -\Delta_{\theta_{g}}(L_{reg})$ \textcolor{black}{(Gradient update)}\; 
            
        }{ load the example ($x_c,y_c,x_{cb}^e,y_{cb}^e$) \;
           $x_c^t$, $y_c^t$ $\gets$ calculated using (\ref{eqn:reg_phi1}) \;
           $L_{seg}^{ohem}$ $\gets$ calculated using  (\ref{eqn:segmentation_loss2_OHEM})\;			
		$\theta_{s} \overset{+}{\gets} -\Delta_{\theta_{s}}(L_{seg}^{ohem})$ \textcolor{black}{(Gradient update)}\; 
        }
		}
	}
	}
		\caption{\label{algo:algorithm}\small{One-shot PACS method.}}
\end{algorithm}

\subsection{Implementation details}
All networks were implemented using Pytorch library and trained on Nvidia GTX V100 with 16 GB memory. The networks were optimized using ADAM algorithm with an initial learning rate of 2e-4 for the first 30 epochs and then decayed to 0 in the next 30 epochs and a batch size of 1. We set $\lambda_{smooth}$=30 experimentally. Eight CLSTM steps were used for both RRN and RSN. GPU memory limitation was addressed using truncated backpropagation through time (TBPTT)\cite{jaeger2002tutorial} after every 4 CLSTM steps. 
\par
The RSN was constructed with 3D Unet with the CLSTM placed on the encoder layers. Each convolutional block was composed of two convolution units, ReLU activation, and max-pooling layer. This resulted in feature sizes of 32,64,128,256, and 512. The RRN extended the Voxelmorph architecture\cite{balakrishnan2019voxelmorph} with CLSTM implemented in the encoder layers. Diffeomorphic deformation was ensured by using a diffeomorphic integration layer\cite{dalca2019unsupervised} following the 3-D flow field output of the CLSTM. 
\textcolor{black}{The last layer of the RRN was composed of a spatial transformation function based on spatial transform networks\cite{jaderberg2015spatial} to convert the feature activations into DVF.} The 3D networks architecture details are in the Supplementary document Table I and II. 

\section{Experiments and Results}
\subsection{Dataset and Experiments: } A retrospective dataset of 369 fully anonymized weekly 4D-CBCT acquired from 65 patients with locally advanced non-small cell lung cancer and treated with intensity modulated radiotherapy using conventional fractionation with a single 4D pCT and up to 6 4D CBCTs acquired weekly during treatment were analyzed. Thirteen out of 65 patients were sourced from an external institution cohort\cite{hugo2017}. Mid phase CTs and CBCTs were analyzed. The scans had an image resolution that ranged from 0.98 to 1.17mm in-plane and 3mm slice thickness. \textcolor{black}{CBCT scans were acquired on a commercial CBCT scanner (On-board Imager$^{TM}$, Varian Medical Systems Inc,) using truebeam with (external: a peak kilovoltage (kVp) of 125kVp, tube current of 50 mAs; internal: 100kVp and 20 mAs) and reconstructed using Ram-Lak filter.}  
\par
The open-source dataset\cite{hugo2017} provided expert delineations. In the internal dataset, the gross tumor volume and the esophagus contours were produced on the pCT and CBCT by an experienced radiation oncologist and these represented the ground truth\cite{jiang2021CBCTMedPhys}. The esophagus contours were outlined on CBCT below the level of cricoid junction to the entrance of the stomach. 
CBCT and pCT scans were rigidly aligned using bony anatomy to bring them in the same spatial coordinates. FOV differences were addressed by resampling CBCT images to the same voxel size as the pCTs and the body mask was extracted through automatic thresholding ($\geq$ 800HU) for soft tissue and the extracted region used as region of interest as done by other prior works\cite{park2017deformable_sift,SIFT_val}.
\par
\textbf{Metrics:} Segmentation was evaluated using the Dice similarity coefficient 
(DSC), surface DSC (sDSC), and Hausdorff distance at 95th percentile (HD95) on \textcolor{black}{testing set}. The tolerance value of 4.38mm for computing sDSC was obtained using two physician segmentations\cite{jiang2021CBCTMedPhys}. \textcolor{black}{Inter-rater accuracy comparisons were done using the DSC metric.} Registration was evaluated using segmentation accuracy, measures of deformation smoothness, namely, standard deviation of the Jacobian determinant and folding fraction ($\left| J_{\phi}\right|$ $\leq$0 ($\%$)) \cite{zhao2019recursive,deVos2019MedIA} computed from 95,551,488 voxels on the test set, and \textcolor{black}{target registration error (TRE) using 3D scale invariant feature transform (SIFT) features\cite{rister2017volumetric_SIFT,park2017deformable_sift} identified on the pCT and CBCT\cite{SIFT_val,park2017deformable_sift}. In the first step, keypoints are located by  applying convolutions with the difference of Gaussians (DoG) function; in the second step, the feature descriptors consisting of 768 dimensional feature are constructed using geometric moment invariants to characterize the keypoints. Correspondences of the detected SIFT features (547 on average) in the \textcolor{black}{pCT} and \textcolor{black}{CBCT} scans were established using random sample consensus followed by manual verification \cite{park2017deformable_sift}. On average, 22 corresponding features were used for TRE computation per image pair.} 
\par
\textbf{Experimental comparisons: } \textcolor{black}{One-shot PACS segmentations} were compared against affine image registration, symmetric diffeomorphic registration (SyN)\cite{avants2008symmetric}, deep learning segmentation only methods 3D Unet\cite{cciccek20163d}, Mask-RCNN\cite{he2017mask}, \textcolor{black}{cascaded segmentation}\cite{cao2021cascaded}, and multiple deep registration based segmentation, Voxelmorph\cite{balakrishnan2019voxelmorph}, recursive cascaded registration using 10 cascades\cite{zhao2019recursive}, R2N2\cite{sandkuhler2019recurrent}, and coupled registration and segmentation network U-ReSNet\cite{estienne2019MICCAI}. The Voxelmorph was regularized using segmentation losses from CBCT segmentations\cite{balakrishnan2019voxelmorph}. Full-shot PACS segmentation and full-shot PACS registration based segmentation were computed to establish upper bounds in accuracy and compare one-shot PACS segmentation against registration-based segmentation.
\par

\textbf{Statistical analysis :\/} \rm \textcolor{black}{Statistical comparisons between one-shot PACS and other methods was done using the DSC metrics computed on the testing sets using pairwise, two-sided Wilcoxon signed rank tests at $95\%$ significance level. The effect of treatment based tumor changes on the longitudinal accuracy of CBCT tumor segmentations was measured using one-way repeated measures ANOVA for DSC and HD95. Only p $< 0.05$ were considered to be significant.}

\textbf{Experiments: }Separate networks are trained for tumor and esophagus segmentation, because tumors, which are abnormal structures are likely to have variable spatial and feature characteristics. Due to GPU limitation, the lung tumor segmentation model was computed from \textcolor{black}{image volumes containing the entire thorax from the apex of lung to the level of diaphragm with a size of 192$\times$192$\times$60 obtained by resizing a volume of interest (VOI) of size 300$\times$300$\times$90. The full extent of the chest was visible on each slice}. The esophagus model was computed from images of size 160$\times$160$\times$80 after resizing a VOI of size 256$\times$256$\times$110 enclosing the entire chest on each slice, \textcolor{black}{starting from the cricoid junction till the entrance of the stomach}. All methods were trained with 3-fold cross-validation using 9,800 VOI pairs obtained from 315 scans. The best model was applied on the independent test set of 54 CBCTs. \textcolor{black}{One shot training was done using a randomly selected CBCT scan in the training set. \textcolor{black}{Separately, robustness of tumor segmentation according to the choice of the one-shot CBCT example was evaluated using tumor location (apex: n = 109; inferior: n = 93; and middle: n = 80) and tumor size (small [$\leq$ 5cc]: n = 54; medium [5cc to 10cc]: n = 80; and large [$>$ 10cc]: n = 146)\cite{jiang2018multiple} as selection criteria. Separate models were trained for each example and tested on a set aside data consisting of 28 apex, 29 inferior, 30 middle or centrally located tumors and  29 small, 28 medium, and 32 large tumors.} Network design and ablation tests were performed to evaluate the impact of various losses, joint vs. two-step training, and the utility of CLSTM on tumor segmentation accuracy.} 
\begin{figure}[htbp]
		\begin{center}
			\includegraphics[width=1.0\columnwidth,scale=1]{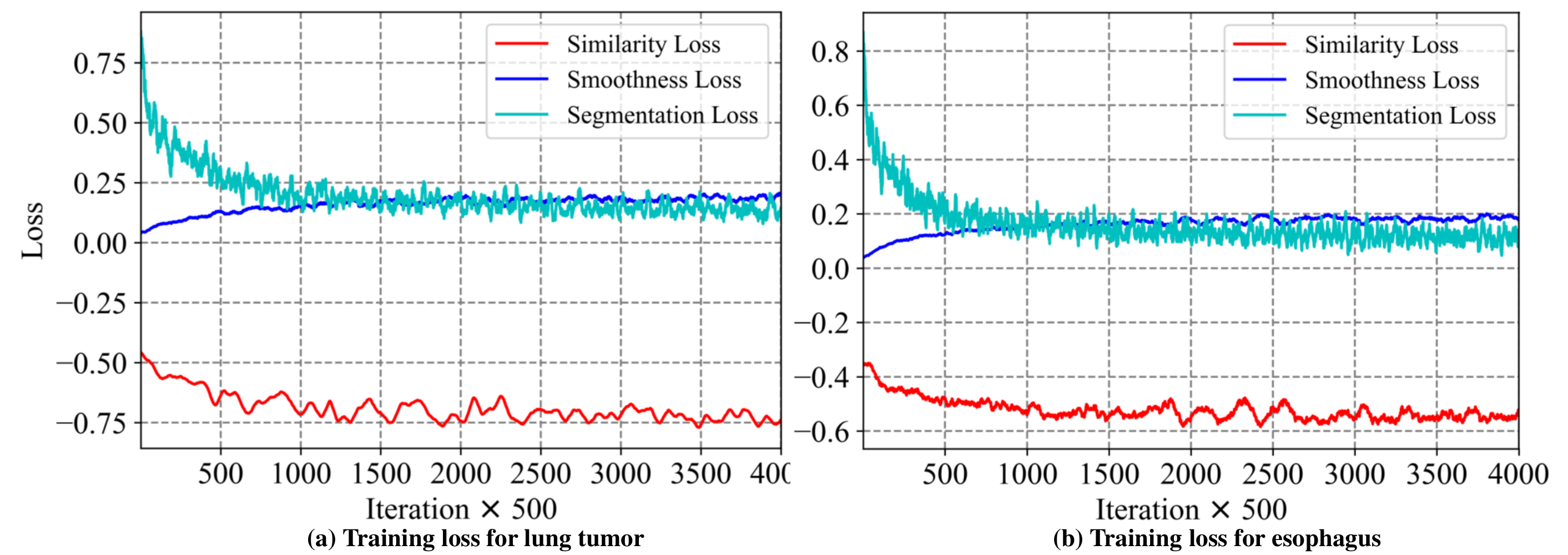}
			\vspace{-0.05cm}\setlength{\belowcaptionskip}{-0.4cm}\setlength{\abovecaptionskip}{0.08cm}\caption{\small \textcolor{black}{Training loss curves for one-shot PACS method.}} \label{fig:train_loss}
		\end{center}
\end{figure}

\begin{figure*}[htbp]
	\begin{center}
		\includegraphics[width=2.0\columnwidth,scale=1]{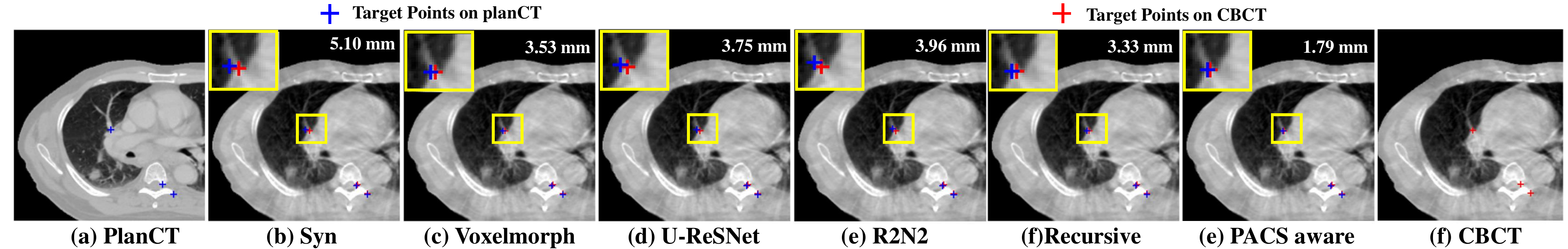}
		\vspace{-0.05cm}\setlength{\belowcaptionskip}{-0.4cm}\setlength{\abovecaptionskip}{0.08cm}\caption{\small \textcolor{black}{SIFT detected targets and the corresponding deformed targets produced by various methods overlaid on CBCT.}}
		\label{fig:TRE_show}
	\end{center}
\end{figure*}

\begin{figure*}[htbp]
		\begin{center}
			\includegraphics[width=2.0\columnwidth,scale=1]{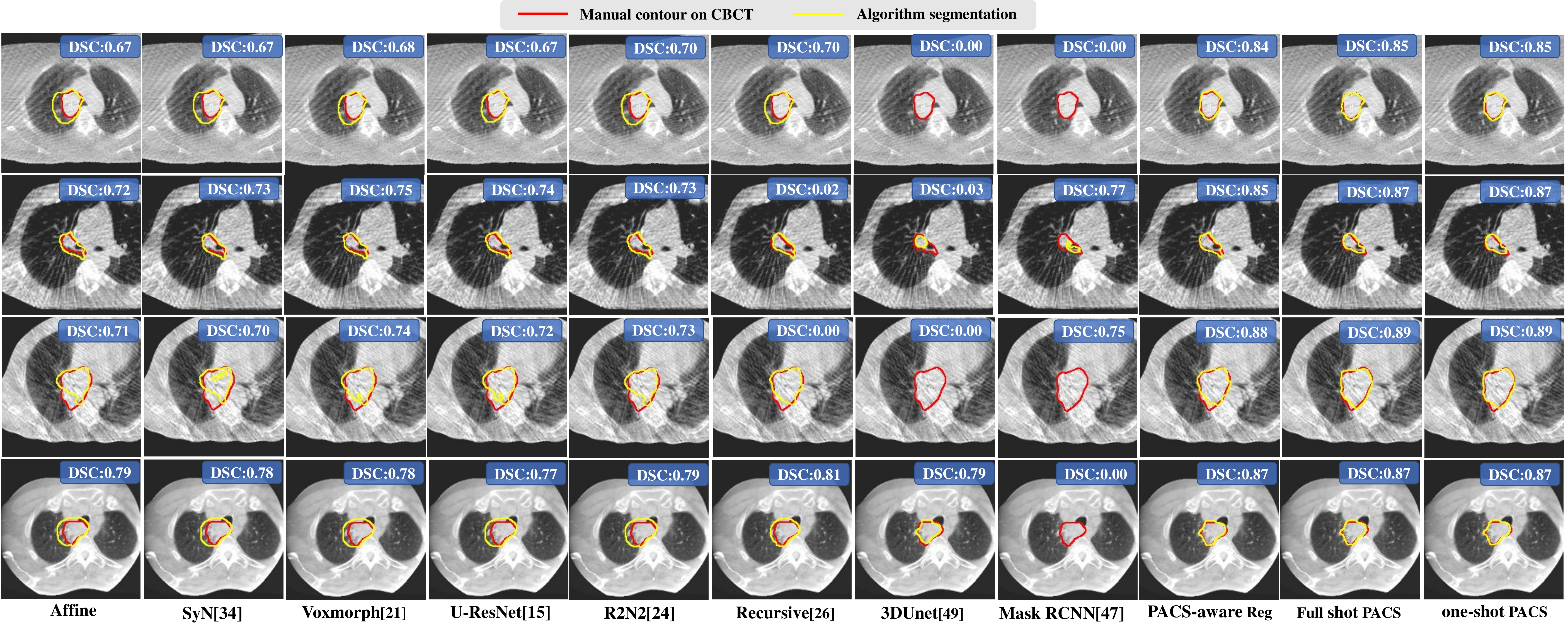}
			\vspace{-0.05cm}\setlength{\belowcaptionskip}{-0.4cm}\setlength{\abovecaptionskip}{0.08cm}\caption{\small Tumor segmentation from CBCT produced by various methods. DSC accuracies for the volume are also shown.} \label{fig:seg_overlay}
		\end{center}
\end{figure*}

\par

\textbf{Network training convergence: \/}\rm \textcolor{black}{Fig.~\ref{fig:train_loss} shows the training loss curves at every 500 iterations for the various losses when training the one-shot PACS network. As shown, the segmentation loss and NCC loss progressively decrease indicating training convergence. Increasing smoothness loss indicates increased image deformation.}

\begin{table} [htb]
\centering{\caption{Registration metrics of various methods.}
\label{tab:jacb} 
	\scriptsize
	\begin{tabular}{|c|c|c|c|c|} 
		\hline 
		{Method}&{SD Jacobian}&{ $\left| J_{\phi}\right|$ $\leq$0 ($\%$)}&{Parameters}&{\textcolor{black}{TRE (mm)}} \\
		\hline
		{SyN\cite{avants2008symmetric}}&{0.04$\pm$0.01}&{0.0022$\pm$0.0066} & {N/A}&{\textcolor{black}{3.94$\pm$1.55}} \\
		\hline
		{Voxelmorph\cite{balakrishnan2019voxelmorph}}&{0.05$\pm$0.01}&{0.0042$\pm$0.011} & {301,411}&{\textcolor{black}{3.13$\pm$1.50}}\\
		\hline
		{Recursive\cite{zhao2019recursive}}&{0.08$\pm$0.02}&{0.013$\pm$0.030} & {42,418,491}&{\textcolor{black}{2.77$\pm$1.53}}\\
		\hline
		{\textcolor{black}{U-ReSNet}\cite{estienne2019u}}& {0.04$\pm$0.01} & {0.021$\pm$0.015} & {4,753,035} & {3.45$\pm$1.62} \\
		\hline
		\textcolor{black}{R2N2}\cite{sandkuhler2019recurrent}  & 0.04$\pm$0.01 & 0.039$\pm$0.012 & 39,183 & 3.25$\pm$1.91 \\
		\hline
		{PACS-aware}&{0.13$\pm$ 0.02}&{0.020$\pm$0.037} & {522,723}&{\textcolor{black}{1.84$\pm$0.76}}\\ 
		\hline
	\end{tabular}
	}
\end{table}

\subsection{Registration smoothness and accuracy}
\textcolor{black}{One-shot PACS produced the lowest TRE of 1.84$\pm$0.76 mm and smooth deformations that were within the accepted range of 1\% of the folding fraction \cite{zhao2019recursive,deVos2019MedIA} (Table.~\ref{tab:jacb}).} \textcolor{black}{It required fewer parameters than recursive \cite{zhao2019recursive}, but more than the R2N2 registration\cite{sandkuhler2019recurrent}. An example slice with corresponding SIFT features from warped pCT produced using different methods overlaid on the CBCT image is shown in Fig.~\ref{fig:TRE_show}}.  

\begin{figure*}[ht]
	\begin{center}
		\includegraphics[width=2.0\columnwidth,scale=1]{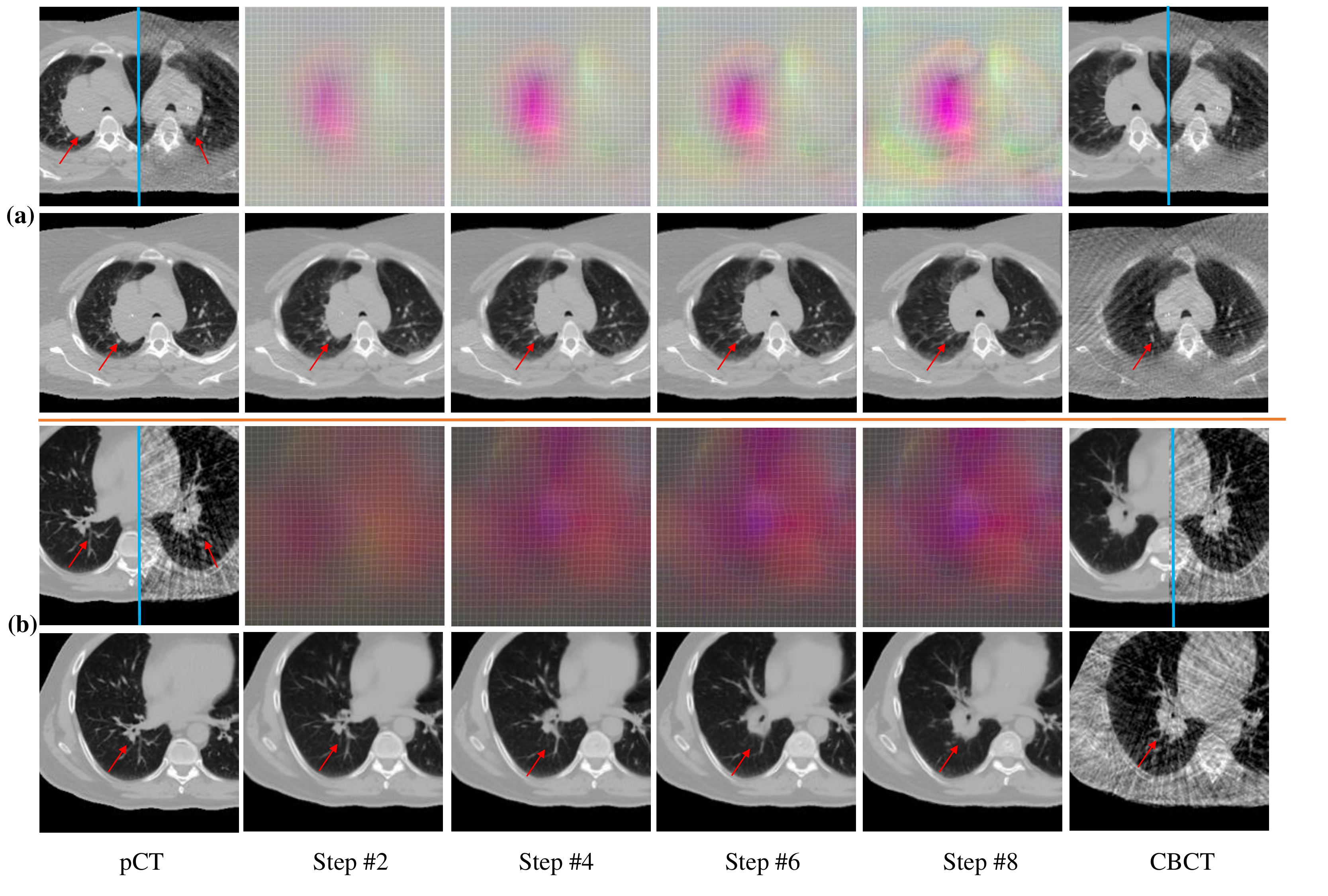}
		\vspace{-0.05cm}\setlength{\belowcaptionskip}{-0.4cm}\setlength{\abovecaptionskip}{0.08cm}\caption{\small Progressive DVFs with warped pCTs (rows 2 , 4) for a shrinking tumor (row 1), and out of plane rotation (row 3). Mirror flipped view of pCT and the CBCT before and after alignment are shown. DVF colors indicate displacements in x (\textcolor{black}{0mm to 10.13mm) (black to red)},y \textcolor{black}{(0mm to 7.76mm) (black to green)}, and z \textcolor{black}{(0mm to 13.50mm) (black to blue)} directions. Red arrow identifies the tumor.}
		\label{fig:img_deform_show}
	\end{center}
\end{figure*}

Fig.~\ref{fig:img_deform_show} shows example registrations with the progressively changing DVFs and the warped pCTs produced by the CLSTM steps. A mirror flipped view of the pCT and it's corresponding CBCT before and after the registration, depicting the qualitative alignment of the images is shown. \textcolor{black}{Brighter colored DVF curves correspond to the large deformations occurring in the regions corresponding to the shrinking tumor as well as the boundary of lung due to respiration differences}. \textcolor{black}{Registration performance for a representative case near descending aorta is shown in Supplementary Fig 6, which shows good alignment.} Additional deformation results are in Supplementary Fig 1.

\begin{table}[t]
				\centering{\caption{Tumor segmentation accuracies produced by various methods. Reg - registration, Seg - segmentation.}
					\label{tab:Seg_Result} 
					
					\scriptsize
					\footnotesize
					
					\begin{tabular}{|c|c|c|c|} 
						\hline 
						
						\hline 
						\multirow{2}{*}{Method}&  \multicolumn{3}{c|}{Testing (\textcolor{black}{Number}=54)}\\ 
						\cline{2-4}
						{    }  & {  DSC  }& {  sDSC} & {  HD95 $mm$  }\\
						\hline 
						
						{Affine Reg} & {0.71$\pm$0.14  }& {0.87$\pm$0.14 }   & {7.31$\pm$3.65  }\\
					\multirow{1}{*}{SyN\cite{avants2008symmetric} } & {0.72$\pm$0.14  }& {0.88$\pm$0.13 }   & {7.03$\pm$3.52  }\\
                    \hline									
\multirow{1}{*}{\textcolor{black}{Voxelmorph}\cite{balakrishnan2019voxelmorph} } & { 0.75$\pm$0.13 }& {0.92$\pm$0.11}   & {5.62$\pm$3.05 } \\
 \multirow{1}{*}{\textcolor{black}{R2N2}\cite{sandkuhler2019recurrent} }& {0.74$\pm$0.13 }& {0.91$\pm$0.10 }   & {6.12$\pm$2.85 } \\
\multirow{1}{*}{\textcolor{black}{U-ReSNet}\cite{estienne2019u} } & {0.73$\pm$0.14  }& {0.90$\pm$0.12 }   & { 6.44$\pm$3.33} \\

					\multirow{1}{*}{Recursive\cite{zhao2019recursive}} & {0.77$\pm$0.11 }& {0.93$\pm$0.08 }   & {5.38$\pm$2.57  }
					\\					
					\cline{1-4}
					{3D Unet\cite{cciccek20163d}} &{ 0.61$\pm$0.15 }& {0.83$\pm$0.15 } & { 16.72$\pm$23.50 }
					\\
{\textcolor{black}{Mask RCNN\cite{he2017mask}}} &{ 0.64$\pm$0.16 }& { 0.82$\pm$0.14} & { 20.53$\pm$23.29 }\\
{\textcolor{black}{Cascaded Net\cite{cao2021cascaded}}} &{ \textcolor{black}{0.63$\pm$0.16}}& { \textcolor{black}{0.81$\pm$0.14}} & { \textcolor{black}{22.61$\pm$23.42} }
					\\					
					\cline{1-4}
					\multirow{1}{*}{PACS-aware Reg } & { 0.81$\pm$0.08  }& { 0.97$\pm$0.05 } & { 4.15$\pm$1.82  }
					\\
                    \multirow{1}{*}{Full-shot PACS seg} & { \textcolor{white}{a}\textbf{0.84$\pm$0.08 } }& {\textcolor{white}{a}\textbf{0.98$\pm$0.04 } } & {\textbf{\textcolor{white}{a}3.33$\pm$2.02  } }
                    \\
					\hline
            		\multirow{1}{*}{One-shot PACS seg} & { \textcolor{white}{a}\textbf{0.83$\pm$0.08 } }& {\textcolor{white}{a}\textbf{0.97$\pm$0.06 } } & {\textbf{\textcolor{white}{a}3.97$\pm$3.06  } }
            		\\
					\hline
					 \cline{1-4}
						
					\end{tabular}
				}
			\end{table}

\subsection{Segmentation accuracy}
\subsubsection{Tumor segmentation}
Table~\ref{tab:Seg_Result} shows the segmentation accuracies produced using various methods. There was no difference in accuracy between the one-shot and full-shot PACS  segmentation (DSC p$=$0.16). One-shot PACS segmentation was significantly more accurate (p$<$0.001) than all other methods, including the full-shot PACS registration-based segmentation (DSC p$=$1.2e-9). Fig.~\ref{fig:seg_overlay} shows segmentations produced by the various methods on randomly selected and representative cases from the external institution dataset. One shot PACS closely approximated expert's segmentations, despite imaging artifacts, indicating feasibility for tumor segmentation.
\begin{figure}[htbp]
		\begin{center}
			\includegraphics[width=0.95\columnwidth,scale=1]{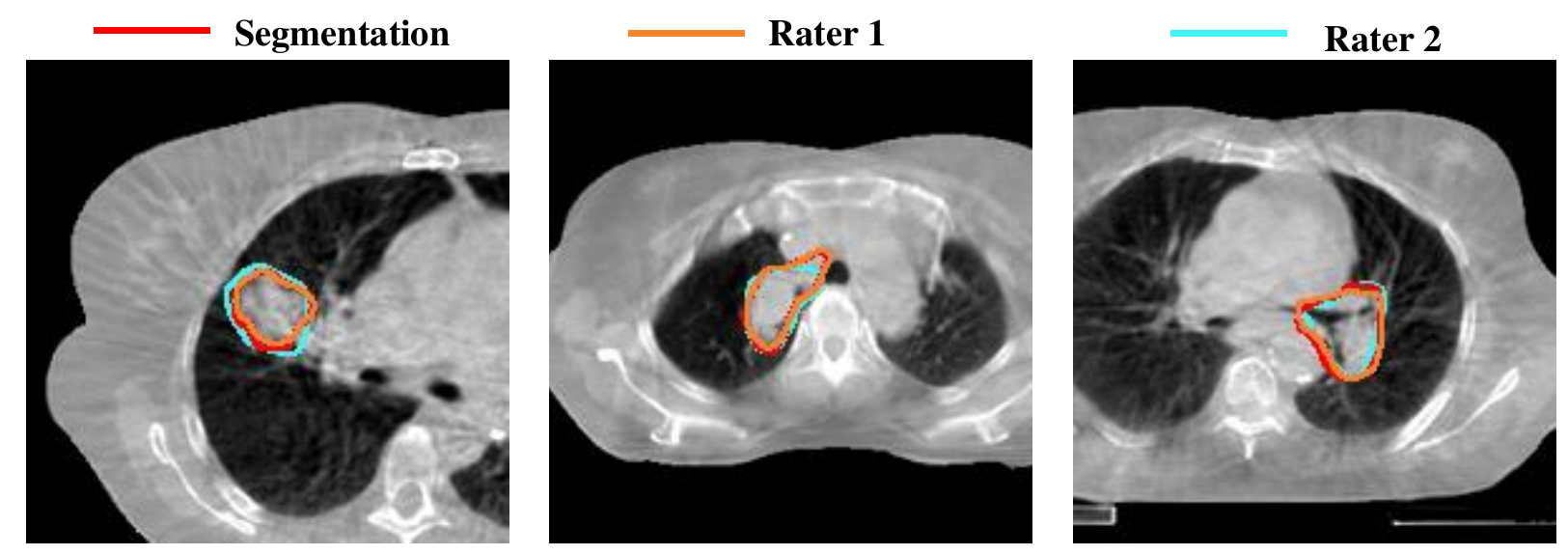}
			\vspace{-0.05cm}\setlength{\belowcaptionskip}{-0.4cm}\setlength{\abovecaptionskip}{0.08cm}\caption{\small \textcolor{black}{One-shot PACS segmentation compared to two raters}.} \label{fig:inter_rater}
		\end{center}
\end{figure}

\textbf{\textcolor{black}{Inter-observer variability:} \/}\rm \textcolor{black}{Robustness to two rater tumor segmentations was measured for 9 patients. One-shot PACS produced a DSC of 0.82$\pm$0.08 and 0.84$\pm$0.09 for raters 1 and 2. The inter-rater DSC was 0.83$\pm$0.06. Fig.~\ref{fig:inter_rater} shows three examples with one-shot PACS and two rater segmentations.}

\subsubsection{Esophagus segmentation} 
\begin{figure*}[tb]
	\begin{center}
		\includegraphics[width=2.0\columnwidth,scale=1]{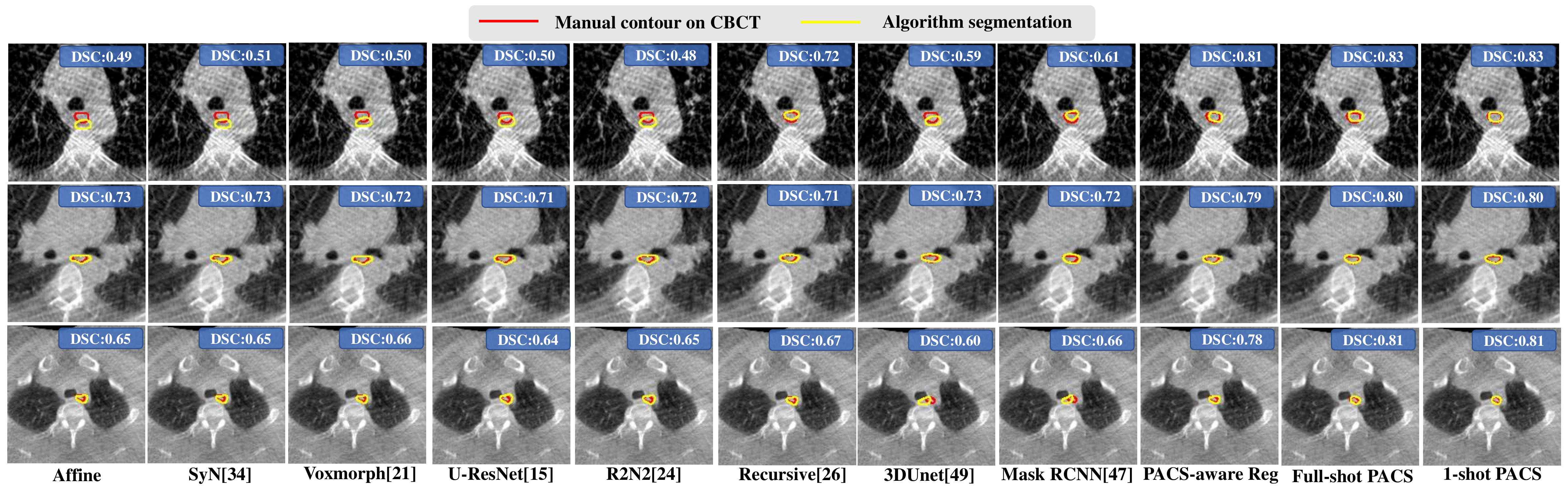}
		\vspace{-0.05cm}\setlength{\belowcaptionskip}{-0.4cm}\setlength{\abovecaptionskip}{0.08cm}\caption{\small Esophagus segmentation from CBCT produced by various methods. Volumetric DSC accuracies are also shown.} \label{fig:seg_overlay_eso }
	\end{center}
\end{figure*}

Table~\ref{tab:Seg_Result_Eso} shows the accuracies for segmenting the esophagus on CBCT images. One-shot PACS was similarly accurate as the full-shot PACS segmentation (DSC p$=$0.07). It was also significantly more accurate (p$<$0.001) than all other methods. Fig. \ref{fig:seg_overlay_eso } shows esophagus segmentation produced by various methods.

\subsection{Longitudinal response assessment}
\textcolor{black}{The mean of maximum HD95 distance per patient from the weekly scans was 4.98mm}. The median and inter-quartile range (IQR) of maximum HD95 from different patients were 4.97 mm and 3.90mm to 6.11mm. \textcolor{black}{Longitudinal accuracy evaluation was done on 30 test patients who had CBCTs from all 6 weeks. The percent slope of DSC accuracy was -0.3\% and HD95 was -8.4\% using HD95 from week 1 to week 6 (Fig.~\ref{fig:weekly_seg_tumor_show}). A one way repeated measures ANOVA with lower-bound corrections determined that CBCT tumor segmentations did not differ between weekly time points (DSC: F(5, 1.35) $=$ 0.0033, p $=$ 0.26; HD95: F(5, 0.56) $=$ 1.56, p $=$ 0.46). There was no significant interaction of tumor location and time on accuracy (DSC: F(5, 0.82) $=$ 0.0002, p $=$ 0.37; HD95: F(5, 1.89) $=$5.22, p $=$ 0.18). These results indicate that the one-shot PACS produced reliable segmentations on weekly CBCT. The one-way repeated measures ANOVA analysis for the esophagus  also did not show a significant effect with time (p $=$ 0.26). The longitudinal accuracy graphs for esophagus are in Supplementary Fig. 8.}   

Fig.~\ref{fig:weekly_seg_tumor_eso} shows the representative example case with volumetric segmentations produced using one-shot PACS and the expert for tumor and esophagus on weekly scans. Our method closely followed the expert delineations. 


\begin{figure}[htbp]
		\begin{center}
			\includegraphics[width=1.0\columnwidth,scale=1]{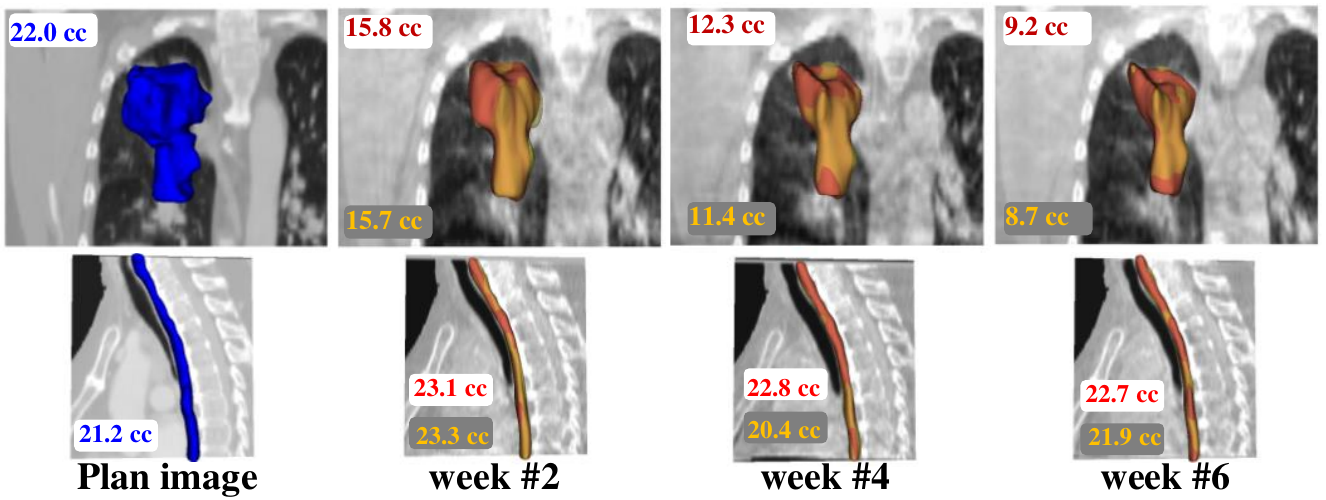}
			\vspace{-0.05cm}\setlength{\belowcaptionskip}{-0.4cm}\setlength{\abovecaptionskip}{0.08cm}\caption{\small Longitudinal segmentation using algorithm (red) and expert (yellow) on weekly CBCT for tumor (top row) and the esophagus (bottom row). The pCT delineation is shown in blue.} \label{fig:weekly_seg_tumor_eso}
		\end{center}
\end{figure}

\begin{figure}[htbp]
		\begin{center}
			\includegraphics[width=1.0\columnwidth,scale=1]{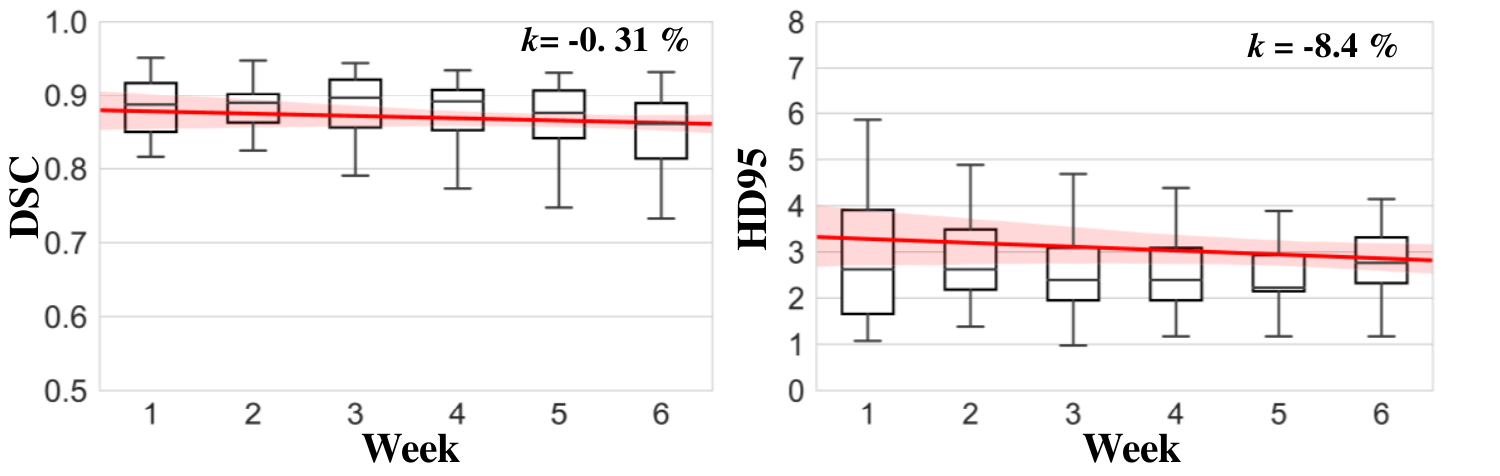}
			\vspace{-0.05cm}\setlength{\belowcaptionskip}{-0.4cm}\setlength{\abovecaptionskip}{0.08cm}\caption{\small \textcolor{black}{Segmentation accuracy at different weeks with percent slope change in accuracy for DSC and HD95 metrics.}} \label{fig:weekly_seg_tumor_show}
		\end{center}
\end{figure}

\begin{figure}[htbp]
		\begin{center}
			\includegraphics[width=1.0\columnwidth,scale=1]{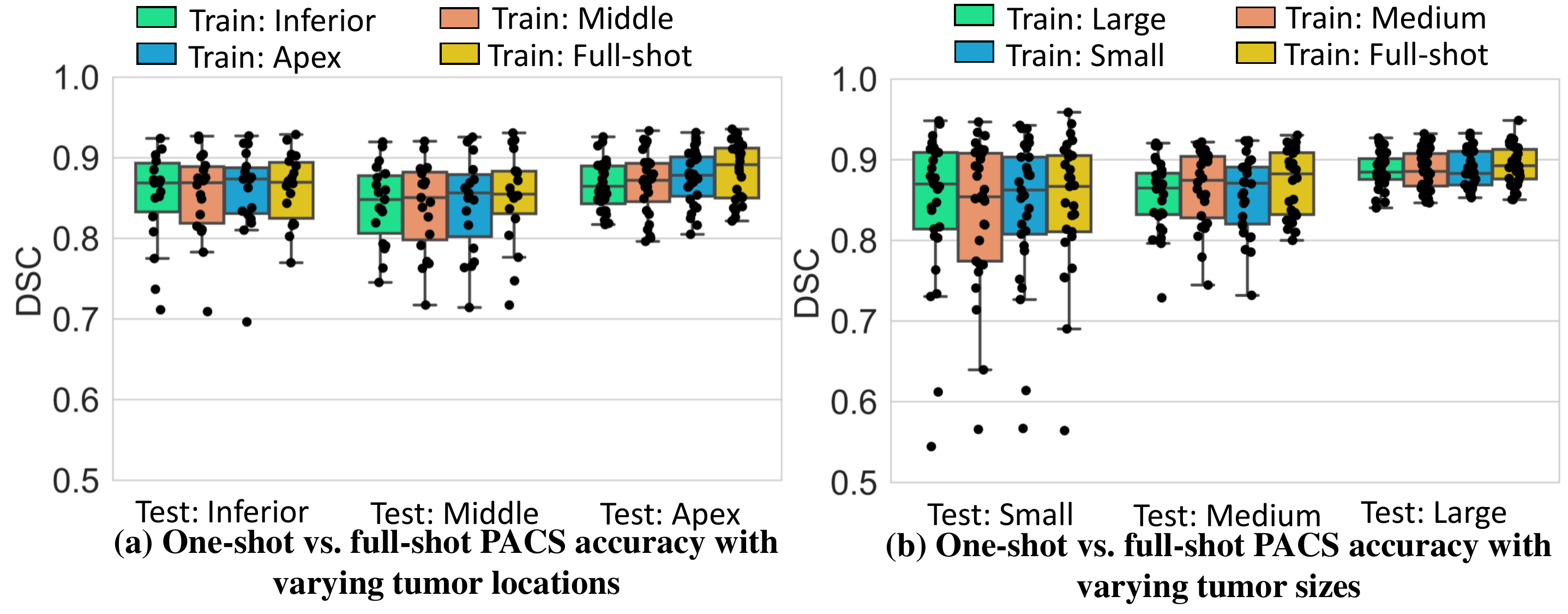}
			\vspace{-0.05cm}\setlength{\belowcaptionskip}{-0.4cm}\setlength{\abovecaptionskip}{0.08cm}\caption{\small \textcolor{black}{Testing set segmentation accuracy with models trained using different one-shot examples by (a) location and (b) size.}} \label{fig:acc_diff_location}
		\end{center}
\end{figure}

\begin{table} [b]
				\centering{\caption{Esophagus segmentation accuracies. Reg - registration, Seg - segmentation. }
					\label{tab:Seg_Result_Eso} 
					\centering
					\small
					\centering
					\begin{tabular}{|c|c|c|c|} 
						\hline 
						
						\hline 
						\multirow{2}{*}{Method}&  \multicolumn{3}{c|}{Testing (\textcolor{black}{Number}=54)}\\ 
						\cline{2-4}
						{    }  & {  DSC  }& {  sDSC} & {  HD95 $mm$  }\\
						\hline 
						
						{Affine Reg} & {0.67$\pm$0.16 }& {0.77$\pm$0.18}   & { 5.11$\pm$3.38}\\
					\multirow{1}{*}{SyN\cite{avants2008symmetric} } & {0.69$\pm$0.17}& {0.79$\pm$0.19}   & {4.96$\pm$3.43}\\
                    \hline									
\multirow{1}{*}{\textcolor{black}{Voxelmorph}\cite{balakrishnan2019voxelmorph} } & { 0.72$\pm$0.15 }& {0.83$\pm$0.17}   & {4.40$\pm$2.89 } \\
 \multirow{1}{*}{\textcolor{black}{R2N2}\cite{sandkuhler2019recurrent} }& {0.73$\pm$0.15 }& {0.84$\pm$0.17 }   & {4.33$\pm$3.14 } \\
\multirow{1}{*}{\textcolor{black}{U-ReSNet}\cite{estienne2019u} } & {0.72$\pm$0.15  }& {0.83$\pm$0.17 }   & { 4.47$\pm$3.20} \\

						\multirow{1}{*}{Recursive\cite{zhao2019recursive}} & {0.73$\pm$0.15}& {0.85$\pm$0.16}   & { 4.11$\pm$2.15 }\\					
						\cline{1-4}
						

						{3D Unet\cite{cciccek20163d}} & {0.57$\pm$0.18  }& { 0.64$\pm$0.17} & { 6.79$\pm$2.87 }\\
{\textcolor{black}{Mask RCNN \cite{he2017mask}}} &{ 0.61$\pm$0.17 }& { 0.68$\pm$0.16} & { 6.67$\pm$2.67 }	\\	
{\textcolor{black}{Cascaded Net\cite{cao2021cascaded}}} &{ \textcolor{black}{0.60$\pm$0.15}}& { \textcolor{black}{0.66$\pm$0.15}} & { \textcolor{black}{7.58$\pm$2.38 }}\\
						\cline{1-4}
						\multirow{1}{*}{PACS-aware reg} & { 0.76$\pm$0.12  }& { 0.88$\pm$0.13 } & { 3.88$
						\pm$2.83 }\\
                        
						\multirow{1}{*}{Full shot PACS seg} & {\textbf{ 0.79$\pm$0.13}}& {\textbf{ 0.91$\pm$0.12}}& {\textbf{3.10$\pm$2.16 }}\\
						\hline

\multirow{1}{*}{One shot PACS seg} & {\textbf{ 0.78$\pm$0.13}}& {\textbf{ 0.90$\pm$0.14}}& {\textbf{3.22$\pm$2.02 }}\\
						\hline
						
						\hline
						
						
					\end{tabular}
				}
\end{table}

\begin{figure}[t]
	\begin{center}
		\includegraphics[width=0.9\columnwidth,scale=1]{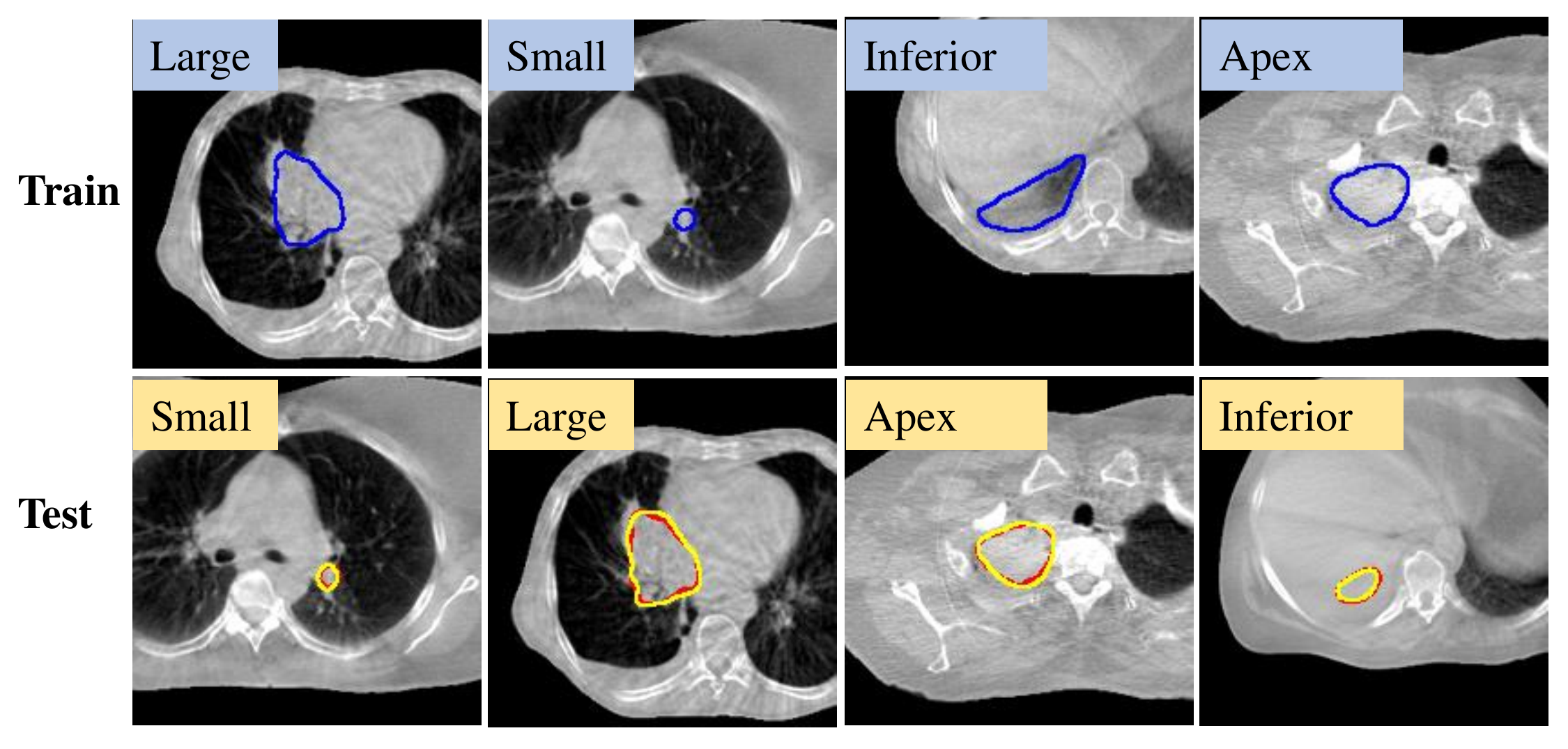}
		\vspace{-0.05cm}\setlength{\belowcaptionskip}{-0.4cm}\setlength{\abovecaptionskip}{0.08cm}\caption{\small Example one-shot PACS segmentor (yellow) results trained with  \textcolor{black}{different sizes and locations}. Red: expert contour.}
		\label{fig:seg_diff_example}
	\end{center}
\end{figure}

\begin{figure*}
	\begin{center}
		\includegraphics[width=2.0\columnwidth,scale=1]{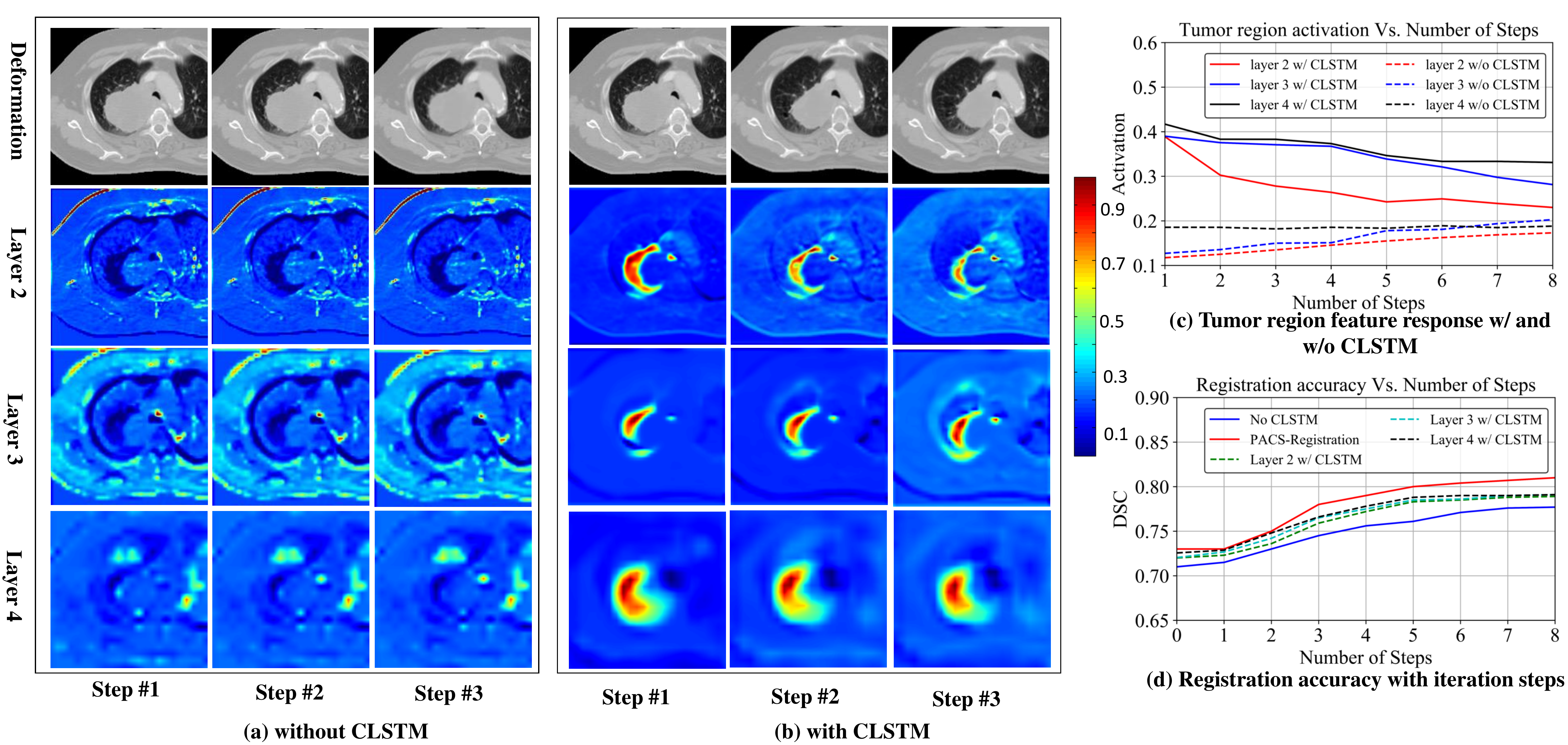}
		\vspace{-0.05cm}\setlength{\belowcaptionskip}{-0.4cm}\setlength{\abovecaptionskip}{0.08cm}\caption{\small \textcolor{black}{(a) Feature activations produced in CNN layers 2, 3, and 4 for steps 1, 2, 3 without CLSTM, and (b) with CLSTM. (c) shows mean feature activations in the layers 2, 3, and 4. (d) shows DSC accuracy with increasing number of recurrent steps.}}
		\label{fig:lstm_vis}
	\end{center}
\end{figure*}
\begin{figure}[htb]
	\begin{center}
		\includegraphics[width=1\columnwidth,scale=1]{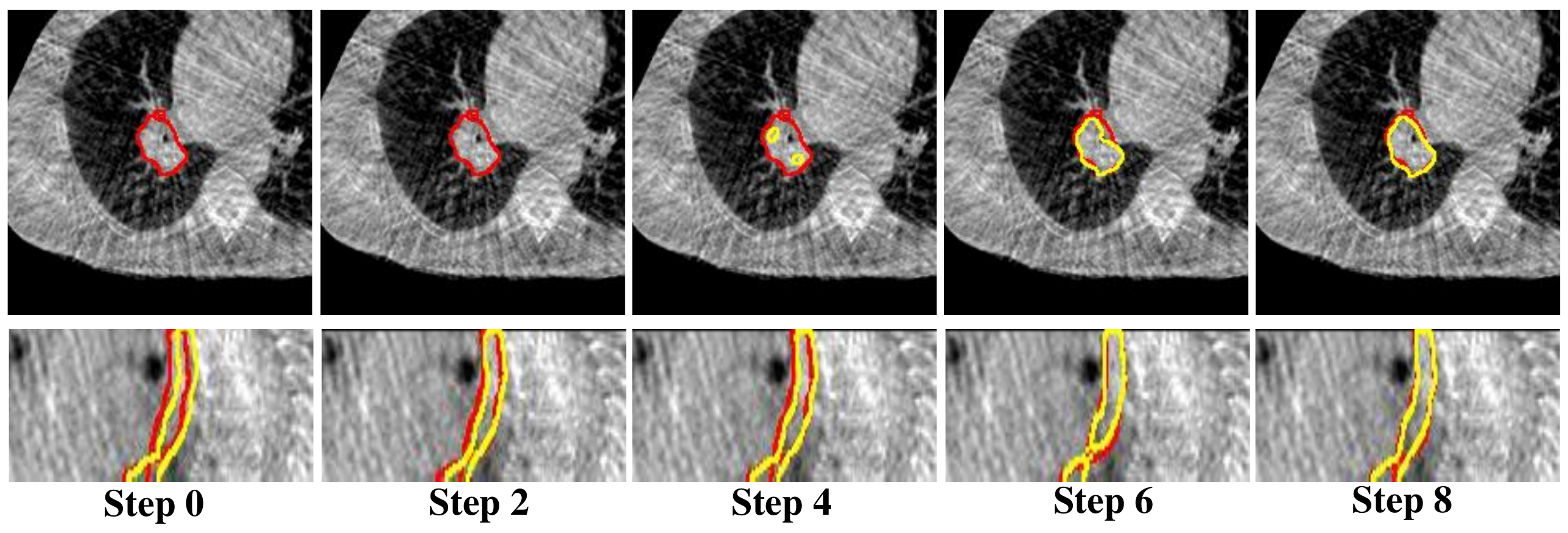}
		\vspace{-0.05cm}\setlength{\belowcaptionskip}{-0.4cm}\setlength{\abovecaptionskip}{0.08cm}\caption{\small Segmentations produced using the one-shot PACS segmentor with increasing number of CLSTM steps for tumor and esophagus. Red is expert, yellow is algorithm contour.}
		\label{fig:seg_vs_iter}
	\end{center}
\end{figure}
\subsection{Network design and ablation experiments} 
\subsubsection{Robustness of one-shot tumor segmentation to selected CBCT training example}
\textcolor{black}{Kruskal-Wallis test showed no difference in the accuracy between one-shot and full shot models for tumor sizes (small: p=0.98; medium: p=0.62; large: p=0.73). Similarly, there was no significant difference between one-shot models trained with examples from different locations and the full-shot model (apex: p=0.29 ; middle: p=0.90; inferior: p=0.99). Summary of mean DSC accuracies as done in\cite{cui2020unified}, produced by the various one-shot models tested on different locations (Fig.~\ref{fig:acc_diff_location}(a)) and sizes (Fig.~\ref{fig:acc_diff_location}(b)), shows similar accuracies for all models. Results for full-shot training is also shown for comparison. Larger variability in accuracy was seen for tumors abutting mediastinum (or middle) and small tumors for all models. Qualitative results on representative cases segmented with one-shot models trained with different tumor locations and sizes show good agreement between algorithm and expert (Fig.~\ref{fig:seg_diff_example}).} 


\subsubsection{Impact of CLSTM in RRN}
\textcolor{black}{We compared the segmentation accuracy when the CLSTM was removed and implemented with convolutional layers, which converted it into a classical dynamic system (CDS) with shared weights for different steps\cite{Goodfellow_book}. Fig.~\ref{fig:lstm_vis} shows the feature activations (steps 1, 2, and 3) produced from layers 2, 3, and 4 (see Supplementary Table I) of the RRN that was trained as a CDS (Fig.~\ref{fig:lstm_vis}(a)) and with CLSTM (Fig.~\ref{fig:lstm_vis}(b)). The step outputs in the case of CLSTM correspond to hidden feature $h_{t}$ of CLSTM, whereas for the CDS corresponds to the feature output after step $t$. Fig.~\ref{fig:lstm_vis} shows alignment of pCT with CBCT image with a centrally located shrinking tumor. Stronger feature activations with a consistent progression in the deformations around the tumor region are seen when using CLSTM (Fig.~\ref{fig:lstm_vis}(b)) compared to the CDS network  (Fig.~\ref{fig:lstm_vis}(a)). Correspondingly, the mean feature activations in all the layers are higher in the CLSTM network (Fig.~\ref{fig:lstm_vis} (c)). The CLSTM network also produced higher accuracy than the CDS network trained without CLSTM (Fig.~\ref{fig:lstm_vis} (d)). Concretely, the CDS network produced an accuracy of 0.78 $\pm$ 0.10, compared to the CLSTM network of 0.81 $\pm$ 0.08}.

\subsubsection{Impact of number of CLSTM steps on accuracy}
We analyzed the accuracy and the computational times with increasing number of recurrent steps from 1 to 12. Segmentation accuracy increased and saturated beyond 8 steps (Supplementary Fig. 4). The computational times increased linearly. Therefore, we chose 8 CLSTM steps for our application. Fig.~\ref{fig:seg_vs_iter} shows the progressive improvement in the tumor and esophagus segmentations for a representative case from the various recurrent steps of the RSN trained using one-shot PACS-aware method. \textcolor{black}{Moreover, one-shot PACS took 6.65 secs for training per iteration and 1.47 secs for testing per image pair.}

\subsubsection{Early vs. intermediate fusion of anatomic context and shape priors into RSN}
The default PACS-aware approach combines progressively warped pCT and delineations as additional input channels with the CBCT image into the individual recurrent units placed in the encoder layers of RSN. We tested whether an intermediate fusion strategy, wherein separate encoders are used to compute features from CBCT and the final warped pCT and its delineation, and combined together at the decoder layer of RSN improved accuracy. The schematic of both methods is depicted in Supplementary Fig. 5. Our results showed that the intermediate fusion was less accurate (DSC of 0.72$\pm$0.15 vs. 0.83$\pm$0.08) than the default early fusion approach. This result indicates that combining the progressively warped pCT and its delineations with CBCT through the recurrent network improves accuracy.  
\subsubsection{Different weights for CLSTM steps}
{\color{black} We studied whether assigning larger weights to segmentation losses from the later CLSTM steps had a greater impact on accuracy. For this purpose, the weights on the CLSTM steps in RSN were linearly increased (w=t$/$(N+1)). This approach had marginal impact on accuracy and resulted in a DSC accuracy of 0.82$\pm$0.10 compared to 0.83$\pm$0.08 for the default method.}

\subsubsection{Ablation experiments}
We analyzed the accuracies when removing the different components of the one-shot PACS-aware network, including (I) shape context prior, (II) the anatomic context, (III) deep supervision of the segmentations produced from the intermediate recurrent steps of the segmentation network. We also measured the accuracy (IV) when CLSTM is removed from the segmentor $s$, (V) when training without the OHEM but with regular cross-entropy loss (Eqn.~\ref{eqn:segmentation_loss2}). The default one-shot PACS segmentor results are shown in (VI). As shown in Table~\ref{tab:Abla_Result}, removing the shape context prior led to a clear lowering of accuracy indicating it's importance for one-shot segmentation training. \textcolor{black}{The very low accuracy when removing shape context is because the reported results are for one-shot segmentor and not the registration-based segmentation.} The shape context was also more relevant than the anatomic context with a significant difference (p $<$ 0.001) in accuracies. Similarly, training without the OHEM loss led to lowering of segmentation accuracy. The segmentations produced using the afore-mentioned training scenarios for a representative  performance is shown in Supplementary document Fig. 7. 
\subsubsection{Influence of adding contour loss}
We measured the accuracy when a contour consistency loss was used to regularize the RRN by minimizing the difference in RRN generated and CBCT segmentation. This experiment was performed in the full-shot PACS mode as CBCT segmentations are needed for training both RRN and RSN. This approach produced a DSC of 0.84$\pm$0.08 from RSN and 0.81$\pm$0.08 from RRN-based segmentation, which is the same as the full-shot method trained without consistency loss, indicating equivalence of the two approaches in terms of accuracy. In contrast, one-shot PACS, which is similarly accurate as the full-shot method does, requires one segmented CBCT example for training.
\subsubsection{Joint training or separate training}
We tested whether  optimizing the registration network first followed by segmentation network, as a two-step optimization, to provide shape and anatomic context to the segmentor improved accuracy over a jointly trained network. The two-step method produced a tumor segmentation accuracy of 0.81$\pm$0.09 DSC compared to 0.83 $\pm$ 0.08 using one-shot PACS \textcolor{black}{ and 0.82 $\pm$ 0.08 compared to 0.84$\pm$0.08 in the full-shot setting}, indicating the multi-tasked approach is  beneficial over two-step optimization.

\begin{table} [htb]
				\centering{\caption{Ablation experiments. DS: Deep supervision; }
					\label{tab:Abla_Result} 
					\centering
					\scriptsize
					\begin{tabular}{|c|c|c|c|c|c|c|c|} 
						\hline 
						
						\hline 
						\multirow{2}{*}{}&\multicolumn{7}{c|}{Tumor Segmentation}\\
						\cline{2-8} 
						{}&{pCT}&{Shape}&{CBCT}&{DS}&{CLSTM Seg}&{OHEM}&{DSC}\\ 
						\hline 
 
                    {I}&{$\checkmark$}&{$\times$}&{$\checkmark$}&{$\checkmark$}&{$\checkmark$}&{$\checkmark$}&{0.02$\pm$0.00}\\
                    \hline {II}&{$\times$}&{$\checkmark$}&{$\checkmark$}&{$\checkmark$}&{$\checkmark$}&{$\checkmark$}&{0.79$\pm$0.11}\\ 		\hline 	
 	        {III}&{$\checkmark$}&{$\checkmark$}&{$\checkmark$}&{$\times$}&{$\checkmark$}&{$\checkmark$}&{0.80$\pm$0.10}\\ 
                       \hline 
 {IV}&{$\checkmark$}&{$\checkmark$}&{$\checkmark$}&{$\checkmark$}&{$\times$}&{$\checkmark$}&{0.81$\pm$0.11}\\ 	\hline     
  {V}&{$\checkmark$}&{$\checkmark$}&{$\checkmark$}&{$\checkmark$}&{$\checkmark$}&{$\times$}&{0.80$\pm$0.11}\\ 	\hline   
 {VI}&{$\checkmark$}&{$\checkmark$}&{$\checkmark$}&{$\checkmark$}&{$\checkmark$}&{$\checkmark$}&{\textbf{0.83$\pm$0.08}}\\ 	
                        \hline 
                        
                        \hline

					\end{tabular}
				}
			\end{table}

\section{Discussion}
We introduced a one-shot recurrent and joint registration-segmentation approach to longitudinally segment \textcolor{black}{thoracic} CBCT scans with large intra-thoracic changes occurring during radiotherapy. Our approach, which incorporates patient-specific anatomic context from higher contrast pCT and shape prior from delineated contours on pCT produced more accurate segmentations than multiple methods. \textcolor{black}{Subset analysis showed that our approach was similarly accurate as two raters indicating feasibility of our approach to reduce inter-rater variability in CBCT segmentations. The shape context as well as the anatomic context prior were essential to improving segmentation network's accuracy in the one-shot setting as shown in the ablation experiments. Our approach was more accurate than cross-modality distillation\cite{jiang2021CBCTMedPhys}, which incorporated MRI information for improving CBCT segmentation (DSC of 0.83 $\pm$ 0.08 using one-shot PACS vs. 0.73 $\pm$ 0.10 using MRI-based distillation) on the same dataset, underscoring the importance of spatial and anatomic priors for segmentation. Our method was similarly accurate as  full-shot training for tumors, and robust to the chosen one-shot training example by location and size. \textcolor{black}{However, larger variation in accuracies were observed for small tumors and centrally located tumors for all models including the full-shot model. Previously, we showed lowering of accuracy for centrally located tumors and smaller sized lung tumors\cite{jiang2018multiple} with standard CT scans}. Addition of contour consistency loss in the registration did not improve accuracy, indicating that our one-shot PACS method is a reasonable alternative to full-shot training when large number of segmented examples are unavailable for training.}
\par
\textcolor{black}{Furthermore, combining the RSN with RRN was significantly more accurate than RRN propagated segmentations, confirming prior findings\cite{estienne2019u,ElmahdyMedPhys2019,beljaardsCrossStitch2020,xu2019deepatlas} that multi-tasked methods are more accurate than registration-based segmentation. We also found that the multi-tasked approach was more accurate than a two-step optimization.}  
\par
\textcolor{black}{Our approach handles large anatomical and appearance changes to diseased and healthy tissues during treatment by computing progressive deformations as a sequence by using a 3D CLSTM. CLSTM, which was introduced to model the sequential dynamics of 2D images\cite{shi2015convolutional}, uses convolutions to compute a dense flow, which adds flexibility compared to parametric LSTM methods. We found that our approach was significantly more accurate than R2N2\cite{sandkuhler2019recurrent}, even when it also computes progressive deformations. R2N2\cite{sandkuhler2019recurrent} employs gated recurrent units with the local deformations computed using Gaussian basis functions, which was insufficient to handle the large anatomic changes common in longitudinal CBCT. Analysis of the recurrent component of our network by replacing the CLSTM with a standard classical system showed that CLSTM produced stronger and consistently progressing activations in local regions (e.g. tumors) undergoing large deformations. Also, the accuracy of the network without CLSTM was similar to the recursive\cite{zhao2019recursive}, but the architectures are different. Our network uses a shared feature weights in all steps, whereas a recursive cascade\cite{zhao2019recursive} method uses different models trained jointly for the cascade steps. Finally, as shown in the ablation experiments, inclusion of CLSTM in the segmentation allowed the network to use progressively warped pCT (anatomic context) and pCT delineation (shape context) to improve accuracy further.}
\par
\textcolor{black}{Finally, tumor segmentation using our method showed no significant changes in accuracy with treatment time (due to tumor shrinkage) or tumor location, indicating robustness of the approach for longitudinal response assessment.}
\par
\textcolor{black}{We also evaluated our approach for esophagus segmentation. Our approach can be extended to simultaneously segment multiple organs by feeding the multi-channel organ probability map as shape prior and image as contextual prior to produce multi-channel output for segmentation\cite{jiang2020psigan}}. 
\par
\textcolor{black}{CBCT images also have much lower FOV compared to the corresponding pCT, exacerbating the problem of robust registration. One prior approach by Zhou et.al\cite{ZhouMIA2021} explicitly handled this issue by performing random crops of the pCT images for improving alignment. We like others\cite{park2017deformable_sift,SIFT_val} handled this issue through pre-processing using extraction of chest region and resampling of CBCT and pCT images.}
\par
\textcolor{black}{Our approach has the following limitations. We did not address the issue of motion averaging for precisely defining the gross tumor margin by aligning with all phases of CBCT acquired in a breathing cycle because the goal was longitudinal response assessment. Although our approach showed feasibility to segment tumors with similar variability as two raters, artifacts are not explicitly handled. Accuracy could be improved further by using \textcolor{black}{SIFT features computed from gradients of the deep feature}\cite{Kearney2018} or surface points\cite{ZhouMIA2021} for registration. Our approach of computing dense flow fields may be adversely impacted especially for abdominal organs which have uniform density internally. In such cases, surface points as used in\cite{ZhouMIA2021} or a MRI-based feature distillation\cite{jiang2021CBCTMedPhys,Fu2020MedPhys} could potentially be incorporated with the recurrent network formulation.} Nonetheless, to our best knowledge, ours is the first to handle longitudinal segmentation of hard to segment lung tumors undergoing radiographic appearance and size changes from during treatment CBCTs. 
\section{Conclusion}

We introduced a one shot patient-specific anatomic context and shape prior aware multi-modal recurrent registration-segmentation network for segmenting on treatment CBCTs. Our approach showed promising longitudinal segmentation performance for lung tumors undergoing treatment and the esophagus on one internal and one external institution dataset. 
{\tiny
	\bibliographystyle{IEEEtran}
}
\bibliography{bib}

\begin{thebibliography}{10}
\providecommand{\url}[1]{#1}
\csname url@samestyle\endcsname
\providecommand{\newblock}{\relax}
\providecommand{\bibinfo}[2]{#2}
\providecommand{\BIBentrySTDinterwordspacing}{\spaceskip=0pt\relax}
\providecommand{\BIBentryALTinterwordstretchfactor}{4}
\providecommand{\BIBentryALTinterwordspacing}{\spaceskip=\fontdimen2\font plus
\BIBentryALTinterwordstretchfactor\fontdimen3\font minus
  \fontdimen4\font\relax}
\providecommand{\BIBforeignlanguage}[2]{{%
\expandafter\ifx\csname l@#1\endcsname\relax
\typeout{** WARNING: IEEEtran.bst: No hyphenation pattern has been}%
\typeout{** loaded for the language `#1'. Using the pattern for}%
\typeout{** the default language instead.}%
\else
\language=\csname l@#1\endcsname
\fi
#2}}
\providecommand{\BIBdecl}{\relax}
\BIBdecl

\bibitem{sonke2019}
J.-J. Sonke, M.~Aznar, and C.~Rasch, ``Adaptive radiotherapy for anatomical
  changes,'' \emph{Semin Radiat Oncol}, vol.~29, no.~3, pp. 245--257, 2019.

\bibitem{wangRadOnc2013}
J.~Wang, J.~Li, W.~Wang, H.~Qi, Z.~Ma, Y.~Zhang \emph{et~al.}, ``Detection of
  interfraction displacement and volume variance during radiotherapy of primary
  thoracic esophageal cancer based on repeated four-dimensional \textsc{CT}
  scans,'' \emph{Radiat Oncol}, vol.~8, no. 224, 2013.

\bibitem{kuckertz2020}
S.~Kuckertz, N.~Papenberg, J.~Honegger, T.~Morgas, B.~Haas, and S.~Heldmann,
  ``Learning deformable image registration with structure guidance constraints
  for adaptive radiotherapy,'' in \emph{Biomedical Image Registration}, 2020,
  pp. 44--53.

\bibitem{Fu2020MedPhys}
Y.~Fu, Y.~Lei, T.~Wang, S.~Tian, P.~Patel, A.~Jani, W.~Curran, T.~Liu, and
  X.~Yang, ``Pelvic multi-organ segmentation on cone-beam \textsc{CT} for
  prostate adaptive radiotherapy,'' \emph{Med Phys}, vol.~47, no.~8, pp.
  3415--3422, 2020.

\bibitem{jiang2021CBCTMedPhys}
J.~{Jiang}, S.~{Riyahi Alam}, I.~{Chen}, P.~{Zhang}, A.~{Rimner}, J.~O.
  {Deasy}, and H.~{Veeraraghavan}, ``{Deep cross-modality (MR-CT) educed
  distillation learning for cone beam CT lung tumor segmentation},'' \emph{Med
  Phys}, 2021.

\bibitem{Altorjai2012CBCT}
G.~Altorjai, I.~Fotina, C.~Lütgendorf-Caucig, M.~Stock, R.~Pötter, D.~Georg,
  and K.~Dieckmann, ``Cone-beam \textsc{CT}-based delineation of stereotactic
  lung targets: The influence of image modality and target size on
  interobserver variability,'' \emph{Int. J Radiat Oncol Bio Phys}, vol.~82,
  no.~2, pp. e265--e272, 2012.

\bibitem{dong2017Neurocomputing}
P.~Dong, L.~Wang, W.~Lin, D.~Shen, and G.~Wu, ``Scalable joint segmentation and
  registration framework for infant brain images,'' \emph{Neurocomputing}, vol.
  229, pp. 54--62, 2017.

\bibitem{zhaoCVPR2019}
A.~{Zhao}, G.~{Balakrishnan}, F.~{Durand}, J.~V. {Guttag}, and A.~V. {Dalca},
  ``Data augmentation using learned transformations for one-shot medical image
  segmentation,'' in \emph{CVPR}, 2019, pp. 8535--8545.

\bibitem{he2020deep}
Y.~He, T.~Li, G.~Yang, Y.~Kong, Y.~Chen, H.~Shu, J.-L. Coatrieux, J.-L.
  Dillenseger, and S.~Li, ``Deep complementary joint model for complex scene
  registration and few-shot segmentation on medical images,'' in \emph{ECCV},
  vol.~1, 2020.

\bibitem{wang2020lt}
S.~Wang, S.~Cao, D.~Wei, R.~Wang, K.~Ma, L.~Wang, D.~Meng, and Y.~Zheng,
  ``\textsc{LT-N}et: Label transfer by learning reversible voxel-wise
  correspondence for one-shot medical image segmentation,'' in \emph{CVPR},
  2020, pp. 9162--9171.

\bibitem{jia2019MICCAI}
X.~Jia, S.~Wang, X.~Liang, A.~Balagopal, D.~Nguyen \emph{et~al.}, ``Cone-beam
  computed tomography (cbct) segmentation by adversarial learning domain
  adaptation,'' in \emph{MICCAI}, 2019, pp. 567--575.

\bibitem{ZhouMIA2021}
B.~Zhou, Z.~Augenfeld, J.~Chapiro, S.~K. Zhou, C.~Liu, and J.~S. Duncan,
  ``Anatomy-guided multimodal registration by learning segmentation without
  ground truth: Application to intraprocedural \textsc{CBCT/MR} liver
  segmentation and registration,'' \emph{Medical Image Anal.}, vol.~71, p.
  102041, 2021.

\bibitem{xu2019deepatlas}
Z.~Xu and M.~Niethammer, ``Deepatlas: Joint semi-supervised learning of image
  registration and segmentation,'' in \emph{MICCAI}, 2019, pp. 420--429.

\bibitem{estienne2019u}
T.~Estienne, M.~Vakalopoulou, S.~Christodoulidis, E.~Battistela, M.~Lerousseau,
  A.~Carre, G.~Klausner, R.~Sun, C.~Robert, S.~Mougiakakou \emph{et~al.},
  ``U-resnet: Ultimate coupling of registration and segmentation with deep
  nets,'' in \emph{MICCAI}, 2019, pp. 310--319.

\bibitem{beljaardsCrossStitch2020}
L.~Beljaards, M.~S. Elmahdy, F.~Verbeek, and M.~Staring, ``A cross-stitch
  architecture for joint registration and segmentation in adaptive
  radiotherapy,'' in \emph{Med. Imaging with Deep Learning}, 2020, pp. 62--74.

\bibitem{cui2020unified}
H.~Cui, D.~Wei, K.~Ma, S.~Gu, and Y.~Zheng, ``A unified framework for
  generalized low-shot medical image segmentation with scarce data,''
  \emph{IEEE Trans on Med Imaging}, vol.~14, no.~8, 2020.

\bibitem{Foote2019IPMI}
M.~D. Foote, B.~E. Zimmerman, A.~Sawant, and S.~C. Joshi, ``Real-time 2d-3d
  deformable registration with deep learning and application to lung
  radiotherapy targeting,'' in \emph{IPMI}, vol. 11492, 2019, pp. 265--276.

\bibitem{Kearney2018}
V.~Kearney, S.~Haaf, A.~Sudhyadhom, G.~Valdes, and T.~Soldberg, ``An
  unsupervised convolutional neural network-based algorithm for deformable
  image registration,'' \emph{Phys Med Biol}, vol.~63, no.~18, p. 185017, 2018.

\bibitem{Chitphakdithai2010MICCAI}
N.~Chitphakdithai and J.~S. Duncan, ``Non-rigid registration with missing
  correspondences in preoperative and postresection brain images,'' in
  \emph{MICCAI}, 2010, pp. 367--374.

\bibitem{fu2020cone}
Y.~Fu, Y.~Lei, Y.~Liu, T.~Wang, W.~J. Curran, T.~Liu, P.~Patel, and X.~Yang,
  ``Cone-beam computed tomography (\textsc{CBCT}) and \textsc{CT} image
  registration aided by \textsc{CBCT}-based synthetic \textsc{CT},'' in
  \emph{Society of Photo-Optical Instrumentation Engineers (SPIE) Conference
  Series}, vol. 11313, 2020, p. 113132U.

\bibitem{shi2015convolutional}
X.~Shi, Z.~Chen, H.~Wang, D.-Y. Yeung, W.-K. Wong, and W.-c. Woo,
  ``Convolutional \textsc{LSTM} network: A machine learning approach for
  precipitation nowcasting,'' \emph{arXiv preprint arXiv:1506.04214}, 2015.

\bibitem{sandkuhler2019recurrent}
R.~Sandk{\"u}hler, S.~Andermatt, G.~Bauman, S.~Nyilas, C.~Jud, and P.~C.
  Cattin, ``Recurrent registration neural networks for deformable image
  registration,'' \emph{NeurIPS}, vol.~32, pp. 8758--8768, 2019.

\bibitem{wang2021alternative}
S.~Wang, S.~Cao, D.~Wei, C.~Xie, K.~Ma, L.~Wang, D.~Meng, and Y.~Zheng,
  ``Alternative baselines for low-shot 3d medical image segmentation—an atlas
  perspective,'' \emph{AAAI}, 2021.

\bibitem{avants2008symmetric}
B.~B. Avants, C.~L. Epstein, M.~Grossman, and J.~C. Gee, ``Symmetric
  diffeomorphic image registration with cross-correlation: evaluating automated
  labeling of elderly and neurodegenerative brain,'' \emph{Medical Image
  Anal.}, vol.~12, no.~1, pp. 26--41, 2008.

\bibitem{brockMedPhys2017}
K.~Brock, S.~Mutic, T.~McNutt, H.~Li, and M.~Kessler, ``Use of image
  registration and fusion algorithms and techniques in radiotherapy: Report of
  the \textsc{AAPM} radiation therapy committee task group no. 132,'' \emph{Med
  Phys}, vol.~44, no.~7, pp. e43--e76, 2017.

\bibitem{ElmahdyMedPhys2019}
M.~Elmahdy, T.~Jagt, Z.~R.T, Y.~Qiao, R.~Shahzad \emph{et~al.}, ``Robust
  contour propagation using deep learning and image registration for online
  adaptive proton therapy of prostate cancer,'' \emph{Med Phys}, vol.~46,
  no.~8, pp. 3329--3343, 2019.

\bibitem{deVos2019MedIA}
B.~D. {de Vos}, F.~F. Berendsen, M.~A. Viergever, H.~Sokooti, M.~Staring, and
  I.~Išgum, ``A deep learning framework for unsupervised affine and deformable
  image registration,'' \emph{Medical Image Anal.}, vol.~52, pp. 128--143,
  2019.

\bibitem{zhao2019recursive}
S.~Zhao, Y.~Dong, E.~I. Chang, Y.~Xu \emph{et~al.}, ``Recursive cascaded
  networks for unsupervised medical image registration,'' in \emph{CVPR}, 2019,
  pp. 10\,600--10\,610.

\bibitem{lee2019tetris}
M.~C.~H. Lee, K.~Petersen, N.~Pawlowski, B.~Glocker, and M.~Schaap, ``Tetris:
  Template transformer networks for image segmentation with shape priors,''
  \emph{IEEE Trans Med Imaging}, vol.~38, no.~11, pp. 2596--2606, 2019.

\bibitem{balakrishnan2019voxelmorph}
G.~Balakrishnan, A.~Zhao, M.~R. Sabuncu, J.~Guttag, and A.~V. Dalca,
  ``Voxelmorph: a learning framework for deformable medical image
  registration,'' \emph{IEEE Trans. Med Imaging}, vol.~38, no.~8, pp.
  1788--1800, 2019.

\bibitem{zhang2011estimating}
W.~Zhang, P.~Yan, and X.~Li, ``Estimating patient-specific shape prior for
  medical image segmentation,'' in \emph{ISBI}.\hskip 1em plus 0.5em minus
  0.4em\relax IEEE, 2011, pp. 1451--1454.

\bibitem{lee2019image}
M.~C. Lee, O.~Oktay, A.~Schuh, M.~Schaap, and B.~Glocker, ``Image-and-spatial
  transformer networks for structure-guided image registration,'' in
  \emph{MICCAI}, 2019, pp. 337--345.

\bibitem{Goodfellow_book}
I.~Goodfellow, Y.~Bengio, and A.~Courville, \emph{Deep Learning}.\hskip 1em
  plus 0.5em minus 0.4em\relax MIT Press, 2016,
  \url{http://www.deeplearningbook.org}.

\bibitem{wu2016highOHEM}
Z.~Wu, C.~Shen, and v.~d. Hengel, Anton, ``High-performance semantic
  segmentation using very deep fully convolutional networks,'' \emph{arXiv
  preprint arXiv:1604.04339}, 2016.

\bibitem{jaeger2002tutorial}
H.~Jaeger, \emph{Tutorial on training recurrent neural networks, covering BPPT,
  RTRL, EKF and the" echo state network" approach}.\hskip 1em plus 0.5em minus
  0.4em\relax GMD-Forschungszentrum Informationstechnik Bonn, 2002, vol.~5,
  no.~01.

\bibitem{dalca2019unsupervised}
A.~V. Dalca, G.~Balakrishnan, J.~Guttag, and M.~R. Sabuncu, ``Unsupervised
  learning of probabilistic diffeomorphic registration for images and
  surfaces,'' \emph{Medical Image Anal.}, vol.~57, pp. 226--236, 2019.

\bibitem{jaderberg2015spatial}
M.~Jaderberg, K.~Simonyan, A.~Zisserman, and K.~Kavukcuoglu, ``Spatial
  transformer networks,'' \emph{arXiv preprint arXiv:1506.02025}, 2015.

\bibitem{hugo2017}
G.~D. Hugo, E.~Weiss, W.~C. Sleeman, S.~Balik, P.~J. Keall, J.~Lu, and J.~F.
  Williamson, ``A longitudinal four-dimensional computed tomography and cone
  beam computed tomography dataset for image-guided radiation therapy research
  in lung cancer,'' \emph{Med Phys}, vol.~44, p. 762–771, 2017.

\bibitem{park2017deformable_sift}
S.~Park, W.~Plishker, H.~Quon, J.~Wong, R.~Shekhar, and J.~Lee, ``Deformable
  registration of ct and cone-beam ct with local intensity matching,''
  \emph{Physics in Medicine \& Biology}, vol.~62, no.~3, p. 927, 2017.

\bibitem{SIFT_val}
G.~Landry, R.~Nijhuis, G.~Dedes, J.~Handrack, C.~Thieke, G.~Janssens,
  J.~Orban~de Xivry, M.~Reiner, F.~Kamp, J.~J. Wilkens \emph{et~al.},
  ``Investigating \textsc{CT} to \textsc{CBCT} image registration for head and
  neck proton therapy as a tool for daily dose recalculation,'' \emph{Med
  Phys}, vol.~42, no.~3, pp. 1354--1366, 2015.

\bibitem{rister2017volumetric_SIFT}
B.~Rister, M.~A. Horowitz, and D.~L. Rubin, ``Volumetric image registration
  from invariant keypoints,'' \emph{IEEE Trans Image Processing}, vol.~26,
  no.~10, pp. 4900--4910, 2017.

\bibitem{cciccek20163d}
{\"O}.~{\c{C}}i{\c{c}}ek, A.~Abdulkadir, S.~S. Lienkamp, T.~Brox, and
  O.~Ronneberger, ``3d \textsc{U}-\textsc{N}et: learning dense volumetric
  segmentation from sparse annotation,'' in \emph{MICCAI}, 2016, pp. 424--432.

\bibitem{he2017mask}
K.~He, G.~Gkioxari, P.~Doll{\'a}r, and R.~Girshick, ``Mask
  \textsc{R}-\textsc{CNN},'' in \emph{Proc. IEEE ICCV}, 2017, pp. 2961--2969.

\bibitem{cao2021cascaded}
Z.~Cao, B.~Yu, B.~Lei, H.~Ying, X.~Zhang, D.~Z. Chen, and J.~Wu, ``Cascaded
  se-resunet for segmentation of thoracic organs at risk,''
  \emph{Neurocomputing}, vol. 453, pp. 357--368, 2021.

\bibitem{estienne2019MICCAI}
T.~Estienne, M.~Vakalopoulou, S.~Christodoulidis, E.~Battistela, M.~Lerousseau,
  A.~Carre, G.~Klausner, R.~Sun, C.~Robert, S.~Mougiakakou \emph{et~al.},
  ``U-resnet: Ultimate coupling of registration and segmentation with deep
  nets,'' in \emph{International Conference on Medical Image Computing and
  Computer-Assisted Intervention}.\hskip 1em plus 0.5em minus 0.4em\relax
  Springer, 2019, pp. 310--319.

\bibitem{jiang2018multiple}
J.~Jiang, Y.-c. Hu, C.-J. Liu, D.~Halpenny, M.~D. Hellmann, J.~O. Deasy,
  G.~Mageras, and H.~Veeraraghavan, ``Multiple resolution residually connected
  feature streams for automatic lung tumor segmentation from ct images,''
  \emph{IEEE Trans Med Imaging}, vol.~38, no.~1, pp. 134--144, 2018.

\bibitem{jiang2020psigan}
J.~Jiang, Y.-C. Hu, N.~Tyagi, A.~Rimner, N.~Lee, J.~O. Deasy, S.~Berry, and
  H.~Veeraraghavan, ``Psigan: joint probabilistic segmentation and image
  distribution matching for unpaired cross-modality adaptation-based mri
  segmentation,'' \emph{IEEE Trans Med Imaging}, vol.~39, no.~12, pp.
  4071--4084, 2020.

\end{thebibliography}
\end{document}